\documentclass[preprint2]{aastex}
\usepackage{graphicx}
\usepackage{amsmath}
\usepackage{amssymb}
\usepackage{cases}

\shorttitle{Exoplanet Cocktail party}
\shortauthors{Waldmann}

\begin{document}

\title{Of `Cocktail Parties' and Exoplanets}

\author{I. P. Waldmann}
\affil{University College London, Gower Street, WC1E 6BT, UK}
\email{ingo@star.ucl.ac.uk}

\begin{abstract}

The characterisation of ever smaller and fainter extrasolar planets requires an intricate understanding of one's data and the analysis techniques used. Correcting the raw data at the 10$^{-4}$ level of accuracy in flux is one of the central challenges. This can be difficult for instruments that do not feature a calibration plan for such high precision measurements. Here, it is not always obvious how to de-correlate the data using auxiliary information of the instrument and it becomes paramount to know how well one can disentangle instrument systematics from one's data, given nothing but the data itself. 
We propose a non-parametric machine learning algorithm, based on the concept of independent component analysis, to de-convolve the systematic noise and all non-Gaussian signals from the desired astrophysical signal. Such a `blind' signal de-mixing is commonly known as the `Cocktail Party problem' in signal-processing. Given multiple simultaneous observations of the same exoplanetary eclipse, as in the case of spectrophotometry, we show that we can often disentangle systematic noise from the original light curve signal without the use of any complementary information of the instrument. In this paper, we explore these signal extraction techniques using simulated data and two data sets observed with the Hubble-NICMOS instrument. Another important application is the de-correlation of the exoplanetary signal from time-correlated stellar variability. Using data obtained by the Kepler mission we show that the desired signal can be de-convolved from the stellar noise using a single time series spanning several eclipse events. Such non-parametric techniques can provide important confirmations of the existent parametric corrections reported in the literature, and their associated results. Additionally they can substantially improve the precision exoplanetary light curve analysis in the future. 

\end{abstract}

\keywords{methods: data analysis --- methods: statistical --- methods: data analysis ---  techniques: photometric --- techniques: spectroscopic }

\section{Introduction}

The field of transiting extrasolar planets and especially the study of their atmospheres is one of the youngest and most dynamic subjects in current astrophysics. Permanently at the edge of technical feasibility, we have come from the first radial velocity and transit detections \citep{mayor95, marcy98,charbonneau00},  via the first detections of molecular features in hot-Jupiter atmospheres \citep{charbonneau02} to ever more detailed characterisations of multitudes of systems \citep{grillmair08, charbonneau08,snellen10b,bean11, swain08, swain09a,swain09b,tinetti07,tinetti10}. With over 700 exoplanets discovered \citep{schneider11} and over 1200 exoplanetary candidates that await confirmation \citep{borucki11}, the focus of interest shifts from the detection to the characterisation of smaller and smaller targets. The governing factor of this progression is the precision at which we can control our instrument and/or stellar systematics, hence the accuracy with which we can analyse the data. 

To minimise the impact of the systematic noise components, different approaches have been proposed in the past. For space and ground-based observations, eg. Spitzer and Hubble \citep[eg.][]{agol10, beaulieu08, beaulieu11,charbonneau02,charbonneau05,charbonneau08,deming07, gillon10,grillmair08,knutson07a, knutson07b, sing11,snellen10b, bean11b, swain08, swain09a,swain09b,tinetti07,tinetti10}, instrumental systematic noise has been approximated using parametric models, often based on auxiliary information (optical state vectors) such as instrumental temperature, orbital inclination, inter and intra-pixel positions of the point-spread-function. Using optical state vectors to de-correlate one's data is an effective technique  \citep{swain08}.  However for instruments that lack a calibration plan at the precision of 10$^{-4}$, the accuracy of the retrieved optical state vectors (e.g. sensor sampling rates) and the adequacy of the instrument model's definition itself become difficult to determine. Some of the recent controversy over results reported by various teams can be attributed to this circumstance \citep{knutson11,stevenson10, beaulieu11, swain08, gibson11, pont10, hatzes10, brunt10}.

The situation is even further complicated by brightness variability of the planet hosting star, in particular for small, very active M-stars. 
Hence, it is important to work towards an alternative route to quantify or remove systematic noise using non-parametric models that work as independent confirmations of existing results.  \citet{carter09}, \citet{thatte10}, \citet{gibson11} and \citet{waldmann11} have progressed towards non-parametric noise models and signal separation using wavelets, principal component, Gaussian processes and Fourier based analysis. 

In this publication, we propose a new non-parametric method to separate systematic noise from the desired lightcurve signal. Given multiple lightcurves, observed simultaneously using spectrographs for example, we can disentangle our desired astrophysical signal from other time-correlated or non-Gaussian systematic noise sources using un-supervised machine learning algorithms.  We furthermore explore the de-correlation of individual time series spanning several consecutive eclipse events. The importance of this work lies with the fact that no additional knowledge of the system or the star is required, besides the observations themselves. Such non-parametric methods provide a potential new route to de-correlate one's date in the case where the instrument does not feature an adequate calibration plan, the quality of the auxiliary information of the instrument is non-optimal or the host star shows significant activity. Such blind de-convolution techniques provide new insight and powerful validation of the established parametric instrumental models reported in the literature. 

Here we will briefly introduce the more general  theory of blind-source separation and proceed with a description of the algorithm proposed. The efficiency of said algorithm is tested with two synthetic models and two HST/NICMOS data sets available in the public domain and one Kepler (Q2~\&~Q3 data release) time series featuring strong time-correlated stellar variability. Future publications will focus on in-depth discussion of the proposed algorithm to specific data sets.

\section{Background: the Cocktail Party Problem}
\label{independent}

In this section  we will briefly describe the fundamental concepts on which the following analysis is based. Readers familiar with independent component analysis may proceed straight to section~\ref{method}.

Let us consider the analogy of three people talking simultaneously in one room. The speech signals of these people are denoted by $s_{1}(t)$, $s_{2}(t)$ and $s_{3}(t)$. In the same room are three microphones recording the observed signals $x_{1}(t)$, $x_{2}(t)$ and $x_{3}(t)$. The observed signals can be expressed in terms of the original speech signals:

\begin{align}
 \label{intro1}
x_{1}(t) &= a_{11} s_{1}(t) +  a_{12} s_{2}(t) +  a_{13} s_{3}(t) \\\nonumber
x_{2}(t) &= a_{21} s_{1}(t) +  a_{22} s_{2}(t) +  a_{23} s_{3}(t) \\\nonumber
x_{3}(t) &= a_{31} s_{1}(t) +  a_{32} s_{2}(t) +  a_{33} s_{3}(t) \\\nonumber
\end{align}

\noindent instead of assuming $x(t)$ and $s(t)$ to be proper time signals, we drop the time dependence and assume them to be random variables 

\begin{equation}
\label{intro2}
x_{k} = a_{k1} s_{1} + a_{k2} s_{2} + ... + a_{kN}s_{N}, ~~~ \text{for all}~k = 1, ..., N
\end{equation}

\noindent where $a_{kl}$ is a weighting factor (in this case the square of the distance of the speakers to the microphone) and $k,l = 1, ..., N$ are some real coefficients with $N$ being the maximum number of observed signals. The individual time series can also be expressed in terms of vectors where bold lower-case letters denote column vectors and upper-case letter denote matrices:

%

\begin{equation}
\textbf{x} = \textbf{As}
\label{timeseries3}
\end{equation}

\noindent where the rows of $\bf{x}$ comprise the individual time series, $x_{k}$, and similarly $\bf{s}$ is the signal matrix of the individual source signals $s_{l}$. $\bf{A}$ is the 'mixing matrix' consisting of the weights $a_{lk}$. Equation \ref{timeseries3} is also known as the instantaneous mixing model and often referred to as the classical 'Cocktail Party Problem' \citep{icabook, hyvarinen99}. 

The challenge is to estimate the mixing matrix, $\bf{A}$ and its (pseudo)inverse the de-mixing matrix, $\bf{W}$, 

\begin{equation}
\label{demix}
\bf{W} = \bf{A}^{-1}
\end{equation}

\noindent given the observations contained in $\bf{x}$ without any additional prior knowledge of either $\bf{A}$ or $\bf{s}$, or for some methods without restrictions of $\bf{A}$ \& $\bf{s}$.  

Several algorithms have been proposed to find the linear transformation of equation~\ref{timeseries3}. Amongst these are principal component analysis (PCA) \citep{pearson01, manly94, pcabook, press07, oja92}, factor analysis (FA) \citep{pcabook, harman67}, projection pursuit (PP) \citep{friedman87,huber85} and the more recently developed independent component analysis (ICA) \citep{comon94,hyvarinen99, hyvarinen99b, hyvarinen00, icabook, icabook2, icabook3}.

The underlying differences between PCA and FA on one hand and ICA and PP on the other are the underling assumptions on the probability distribution functions (pdfs) of the signals comprising $\bf{x}$. The former group assumes the signals to follow: 1) a Gaussian distribution whilst the latter assume the signals to be, 2) predominantly non-Gaussian or sparse with specific signatures in the spectral domain \citep[e.g. SMICA][]{delabrouille03}. This results in significant differences in the way we estimate our signal components.

1) If the observed signals comprising $\bf{x}$ follow Gaussian distributions, we can define their statistics by the first and second statistical moments (mean and covariance) only. Algorithms such as PCA and FA find a linear transformation from the correlated observed signals, $\bf{x}$, to a set of uncorrelated signals, $\bf{s}$. In this case, uncorrelatedness is equivalent to mutual independence and the source signals are at their most separated (please see Appendix~\ref{appendix:independent} for a more in-depth discussion on independence). Such a linear transformation is always possible and easily achieved using, for example, single value decompositions (SVD) \citep{pearson01, manly94, pcabook, press07}.  An application of PCA to exoplanetary light curve de-trending is given by \citet{thatte10}.

2) In the case of the observed signals following non-Gaussian distributions, significant information is contained in the higher statistical moments (skew \& kurtosis) and it can be shown that uncorrelated signals (as produced by PCA \& FA) are not necessarily mutually independent and hence not optimally separated from one another. Here uncorrelatedness is a weaker constraint than independence and it can be said that independent signals are always uncorrelated but not vice versa (see appendix~\ref{appendix:independent}). As a consequence using PCA or FA algorithms may only yield a partially separated result for non-Gaussian sources. 

It is important to note that most observed signals, astrophysical or stellar/instrumental noise, are predominantly non-Gaussian by nature. We can also state that most of these signals should be statistically independent from one another (e.g. an exoplanet light curve signal is independent of the systematic noise of the instrument with which it was recorded). These properties have led to a surge in ICA based analysis methods in current cosmology and extra-galactic astronomy. Here ICA is used to separate the cosmic microwave background (CMB) or signatures of distant galaxies from their galactic foregrounds \citep[e.g.][]{stivoli06, maino02, maino07, wang10}. \citet{aumont07} furthermore separates instrumental noise from the desired astrophysical signal. Other applications include data-compression of sparse, large data sets to improve model fitting efficiencies \citep[e.g.][]{lu06, delabrouille03}.

\subsection{ICA in the context of exoplanetary lightcurves}

In this publication, we focus on the application of ICA to exoplanetary lightcurve analysis. Let us consider multiple time series observations of the same exoplanetary eclipse signal either in parallel, by performing spectrophotometry with a spectrograph, or consecutive in time (as explained in section~\ref{kepler}).

Without excluding the most general case, let us focus on a time-resolved spectroscopic measurement of an exoplanetary eclipse. For most observations, the signal recorded is a mixture of astrophysical signal, Gaussian (white) noise and systematic noise components originating from instrumental defects and other sources such as stellar activity and telluric fluctuations. We can therefore write the individual time series as sum of the desired astrophysical signal, $s_a$, systematic (non-Gaussian) noise components, $s_{sn}$, and Gaussian noise, $s_{wn}$. We can now define the underlying linear model of our time series data to be

\begin{align}
x(t) = & a_{1} s_{a}(t) +  a_{2} s_{sn1}(t) + a_{3} s_{sn2}(t)  + ... + s_{wn}(t) 
\label{timeseriesexo}
\end{align}

\noindent or 
\begin{equation}
{x}_{k} = a_{k1}{\bf s}_{a} + \sum_{l_{2}=1}^{N_{sn}} a_{kl_{2}} {\bf s}_{sn,l_{2}} + \sum_{l_{3}=1}^{N_{wn}}  a_{kl_{3}} {\bf s}_{wn,l_{3}}
\label{timeseriesexo2}
\end{equation}

\noindent where $N_{sn}$ and $N_{wn}$ are the number of systematic and white noise components respectively and $N~=~N_{sn}~+~N_{wn}~+~1$ assuming only one component is astrophysical.

\subsection{Demixing signals using ICA}
\label{icadef}

The basic assumptions of ICA are that the elements comprising $\bf{s}$, ${s_{l}}$, are mutually independent random variables with probability densities, $p_{l}(s_{l})$. We further assume that all (or at least one) of the probability densities, $p_{l}(\cdot)$, are non-Gaussian. This non-Gaussianity is key since it allows the de-mixing matrix, $\bf{W}$, to be estimated. From the central limit theorem, we know that a convolution of any arbitrary probability distribution functions (pdfs) that feature a formal mean and variance, asymptotically approaches a Gaussian distribution in the limit of large $N$ \citep{riley02}. In other words, the sum of any two non-Gaussian pdfs (ie. $p_{l}(\cdot)$ and $p_{l+1}(\cdot)$) is more Gaussian than the respective original pdfs. Therefore by maximising the non-Gaussianity  of the individual signals, we maximise their statistical independence. \citep{comon94, hyvarinen99, hyvarinen00, koldovsky06, icabook, icabook2, icabook3}. 

\subsubsection{Information Entropy}

Although several measures of non-Gaussianity exist (we refer the reader to \citet{cichocki02}, \citet{hyvarinen00}, \citet{icabook} and \citet{icabook2} for  detailed summaries), we here use the concept of 'negentropy' \citep{brillouin53}. Negentropy is derived from the basic information-theoretical concept of differential entropy. In information theory, in close analogy to thermodynamics, the entropy of a system is at its maximum when all data points are at its most random. In thermodynamics we measure the distribution of particles, in information theory it is the probability distribution of a random variable. From information theory we can derive the fundamental result that a Gaussian distribution has the largest entropy among all random variables of equal variance and known mean. Hence, by minimising the entropy of a variable, we maximise its non-Gaussianity.
For a random vector $\bf{y}$, with random variables $y_{i}$, $i = 1, ..., n$, the entropy is given by 

\begin{equation}
\text{H}({\bf y}) = - \int p({\bf y}) \text{log}_{2} p({\bf y}) \text{d}{\bf y}
\label{shannon}
\end{equation}

\noindent where H$(\bf{y})$ is the differential or Shannon entropy \citep{shannon48} and $p(\bf{y})$ is the pdf of the random vector $\bf{y}$. H$(\bf{y})$ is at its minimum when $p({\bf y})$ is at its most non-Gaussian. We can now normalise equation~\ref{shannon} to yield the definition of negentropy:

\begin{equation}
\text{J}({\bf y}) = \text{H}({\bf y}_{gauss}) - \text{H}({\bf y})
\label{negentropy}
\end{equation}

\noindent where ${\bf y}_{gauss}$ is a random Gaussian vector with the same covariance matrix as $\bf{y}$. Now $\bf{y}$ is at its most non-Gaussian when $\text{J}({\bf y})$ is at its maximum. It is important to note that negentropy is insensitive to a multiplication by a scalar constant. This has the important consequence of introducing a sign and scaling ambiguity into the retrieved signal components of $\bf{s}$. We will discussed the implications of this limitation in section~\ref{simu}. 

\subsubsection{Contrast functions}
\label{contrast}

In practice it is very difficult to calculate the negentropy of a system and various methods were devised to approximate $\text{J}({\bf y})$. The classic method is to measure the kurtosis of mean-subtracted $\bf{y}$ with unit variance. However, kurtosis is very prone to distortions by outliers in the data and hence lacks the robustness required as measure of negentropy \citep{icabook}. To overcome this limitation, more robust measures of negentropy have been devised.  One can approximate negentropy by equation \ref{negentropy2} \citep{icabook, hyvarinen99, icabook2, icabook3}

\begin{equation}
\text J({\bf y}) \propto (\text{E}[G({\bf y})] - \text{E}[G(\nu)])^{2}
\label{negentropy2}
\end{equation}

\noindent where $\nu$ is a random Gaussian variable with zero mean and unit variance and $G$ is a non-quadratic contrast function chosen to approximate the underlying probability distribution. There are usually three types of contrast functions: $G_{1}$ as general purpose function, $G_{2}$ optimised for super-Gaussian (leptokurtic) distributions and $G_{3}$ optimised for sub-Gaussian (platykurtic) distributions \citep{hyvarinen99,icabook, icabook2}:

\begin{align}
\label{nonlin1}
G_{1}(y) &= \frac{1}{a_{1}} log [\text{cosh}( a_{1} y)] \\\nonumber
G_{2}(y) &= -\text{exp}(-y^{2}/2) \\\nonumber
G_{3}(y) &= \frac{1}{4}y^{4}
\end{align}

\noindent The choice of contrast function is only important if one wants to optimise the performance of the ICA algorithm as it is done for the EFICA \citep{koldovsky06} algorithm where choices of contrast functions are tried iteratively.

\subsubsection{FastICA}
\label{fastica}

Finally, we can maximise the negentropy given in equation \ref{negentropy2} by finding a unit vector $\bf{w}$ that maximises the non-Gaussianity of the projection $y_{i} = \bf{w}^{T}\bf{x}$, so that $\text J(\bf{w}^{T}\bf{x})$ is at its maximum. For the fixed-point FastICA algorithm, this can be achieved by the simple iterations scheme \citep{hyvarinen99,hyvarinen00}:

\begin{enumerate}
\item Choose initial (i.e. random) weight vector $\bf{w}$

\item Increment: ${\bf w}^{+} = \text E [ {\bf x} g({\bf w}^{T}{\bf x})] - \text E [ g^{'} ({\bf w}^{T} {\bf x})] {\bf w}$

\item Normalise: ${\bf w} = {\bf w}^{+}/ || {\bf w} ^{+} || $

\item Repeat steps 2 \& 3 until converged

\end{enumerate}

\noindent where $g$ and $g^{'}$ are the first and second derivatives of the contrast function $G(\cdot)$. The iteration scheme above estimates only one weight vector at a time, for estimating all non-Gaussian components in parallel the iteration scheme and derivates of $G(\cdot)$ are given in appendix \ref{efica} and in the standard literature \citep[e.g.][]{hyvarinen99, icabook, icabook2, koldovsky06}. For a comprehensive summary of other ICA algorithms please refer to \citep{icabook2}. Throughout this paper, we use a variant of the FastICA algorithm. 

\subsubsection{Projection Pursuit and ICA}
\label{projectionpursuit}

Projection Pursuit and Independent Component Analysis are inherently linked as both algorithms try to represent the most non-Gaussian components in an multidimensional data set. In this sense, one can think of ICA as a variant of PP with two major differences: 1) PP only estimates one non-Gaussian component at a time whilst ICA is the multivariate definition of PP and estimates all non-Gaussian components simultaneously, 2) as opposed to ICA, PP does not need an underlying data model and no assumptions about independent components are made. If the ICA model holds, optimizing the ICA non-Gaussianity measures produce independent components; if the model does not hold, then what we get are the projection pursuit directions \citep{hyvarinen00,icabook3}. This is an important point to make with regards to time-consecutive transit observations, which break the underlying ICA assumptions otherwise.

\section{The algorithm}
\label{method}

\begin{figure}
\epsscale{1.0}
\plotone{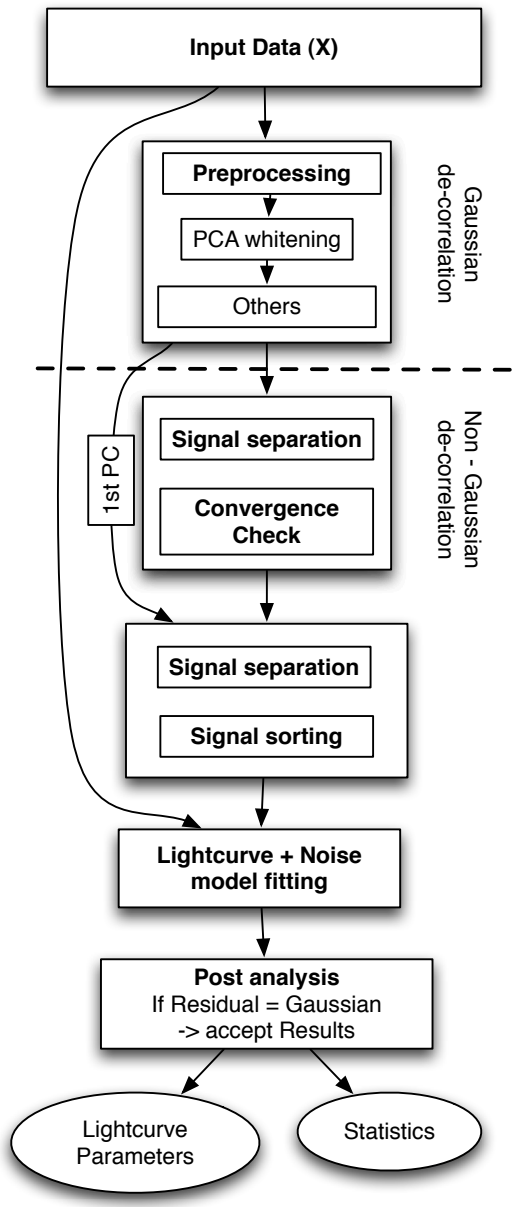}
\caption{Flowchart illustrating the algorithm. The input data is first transformed into an orthogonal set using PCA. The latent signals comprising the input data are then separated using the MULTI-COMBI algorithm which is followed by a signal sorting step. The separated lightcurve and systematic noise components are then fitted to the original data. }
\label{flow}
\end{figure}

Following from the discussion above, we can understand the signal de-mixing to be a two step process: de-correlation of the Gaussian components in the observed data using PCA, followed by the de-correlation of non-Gaussian components using ICA and WASOBI algorithms. The de-correlation of Gaussian components to form a new uncorrelated vectors set can be understood as pre-processing step to the ICA procedure. 
The algorithm proposed here consists of five main parts: 1) Pre-processing of the observed data, 2) Signal separation, 3) Signal reconstruction 4) Lightcurve fitting and  4) Post-analysis. Figure \ref{flow} lays out the individual processing steps of the algorithm. 

\subsection{Signal pre-processing}
\label{pca}
Similarly to section~\ref{independent}, the observed signals can be expressed as $k \times m$  dimensional matrix \textbf{X} where each row constitutes an individual time series, $x_{k}$, with $m$ number of data points. Multiple time series observations are needed to separate the instantaneously mixed non-Gaussian components. 
The process of identifying statistical independent components is greatly simplified if the input signals to any ICA algorithm have previously been whitened (also referred to as sphering). Whitening is essentially a transformation of our input data matrix \textbf{X} into a mean subtracted, $(\bf{X} - \bf{\bar{X}})$, orthogonal matrix $\bf{\tilde{X}}$, where its auto-covariance matrix, $\bf{C}_{\tilde{x}}$, equals the identity matrix, $\bf{C}_{\tilde{x}} = E[\bf{\tilde{X}} \bf{\tilde{X}}^{T}] = I$. The instantaneous mixing model for the whitened data is now given by 

\begin{equation}
\bf{\tilde{X}} = \bf{C_{x}^{-1/2}}(\bf{X} - \bf{\bar{X}}) = \bf{\tilde{A}}\bf{S}
\label{whitening}
\end{equation}

 \noindent where $\bf{C_{x}^{-1/2}}$ is the inverse square root of $\bf{C_{x}}$ and $\bf{\tilde{A}}$ the corresponding mixing matrix of $\bf{\tilde{X}}$. For a more detailed explanation see Appendix. 
 
 This whitening is easily achieved by performing a principal component analysis on the data (see Appendix). This step has two distinct advantages: 
 
 1) It reduces the complexity of the un-whitened mixing matrix, $\bf{A}$, from $n^{2}$ parameters, to $n(n-1)/2$ for a whitened, orthogonal matrix $\bf{\tilde{A}}$ \citep{icabook}. 
 2) Using whitening by principal components, we can reduce the dimensionality of the data-set by only maintaining a sub-set of eigenvectors. This reduces possible redundancies of the components comprising the data and prevents the later to be employed ICA algorithm from over-learning for over-complete sets.    

We also like to note that any type of additional linear signal cleaning or pre-processing step, such as those described by \citet{carter09,waldmann11}, are allowed. Linear data filtering or cleaning can be understood as multiplying equation \ref{timeseries3} from the left with a linear transformation $\bf{B}$ to get: $\bf{B}\bf{X} = \bf{B}\bf{A}\bf{S}$. The underlying data model assumed in this paper is hence not affected. 

\subsection{Signal separation}

After the observed signals have successfully been whitened ($\bf{\tilde{X}}$), we estimate the mixing matrix of the whitened signal, $\bf{\tilde{A}}$, using the MULTI-COMBI algorithm \citep{tichavsky06}. MULTI-COMBI comprises two complimentary algorithms, EFICA  \citep{koldovsky06} and WASOBI \citep{yeredor00}. EFICA, an asymptotically efficient variant of the FastICA algorithm \citep{hyvarinen99}, is designed to separate non-Gaussian, instantaneously mixed signals. WASOBI, on the other hand, is an asymptotically efficient version of the SOBI algorithm \citep{belouchrani97}, and is geared towards separating Gaussian auto-regressive (AR) and time-correlated components. It uses second-order statistics and can be understood to be similar to principal component analysis. The use of both algorithms is necessary since a real life data set will always contain a mixture of both, non-Gaussian and Gaussian AR processes. For a more in-depth discussion of the algorithms employed here, we like to refer the interested reader to the appendices~\ref{efica} \& \ref{wasobi} and the original publications. 

The EFICA and WASOBI algorithm can be shown to be asymptotically efficient, i.e. the estimators approach the Cram\'er-Rao lower bound \citep{davison09}. In other words, the algorithms employed here can be shown to converge to the correct solution given the original source signals and in the limit of $N \rightarrow \infty$ iterations. In reality the number of iterations is finite and and imperfect convergence results in traces of other sources to remain in the individual signals comprising $\bf{S}$.  We can hence state that equation \ref{demix} becomes 

\begin{equation}
\bf{W} \simeq \bf{A}^{-1}
\label{demix2}
\end{equation}

\noindent A measure of this error is the deviation of $\bf{WA}$ (or $\bf{\tilde{W}\tilde{A}}$ for the whitened case) from the unity matrix by inspecting the variance of its elements \citep{koldovsky06,icabook}. 
This leads to the concept of the interference-over-signal ratio (ISR) matrix. The ISR is the standard measure in signal processing of how well a given signal has been transmitted or de-convolved from a mixture of signals. It can be understood as the inverse of the signal-to-noise ratio (SNR). The higher the ISR for a specific signal, the less well has it been separated from the original mixture. For a real case example, $\bf{A}$ is unknown and the ISR needs to be estimated. An analytic approximation to the ISRs for the EFICA and WASOBI algorithms are found in the appendices~\ref{efica} \& \ref{wasobi}. 

Finally, we check the stability of the signal separation by perturbing the input matrix $\bf{\tilde{X}}$ by a random and known matrix $\bf{P}$ to give 

\begin{equation}
\label{perturb}
\bf{\tilde{X}}_{2} = \bf{P} \bf{\tilde{X}} = \bf{P}\bf{\tilde{A}}\bf{S}
\end{equation}

We now re-run the MULTI-COMBI procedure using $\bf{\tilde{X}}_{2}$ as input and estimate $\bf{P}\bf{\tilde{A}}$. Knowing $\bf{P}$ we can work backwards to obtain $\bf{\tilde{A}}$ as we are dealing with a linear transformation. This step is repeated several times to check the convergence of the algorithm by inspecting the variation on the mean ISR values of each separation attempt and the variations in consecutive estimations of $\bf{\tilde{A}}$ directly. 

\subsection{Signal reconstruction}
\label{nmodel}

Once  the mixing matrix, $\bf{\tilde{A}}$ is estimated, we need to identify which signals are astrophysical, which ones are white and which are systematic noise. This is done in a two step process: 

1) We construct the estimated signal matrix, $\bf{\hat{S}}$, and for its individual components $\bf{\hat{s}}_{l}$ compute the Pearson correlation coefficient between $\bf{\hat{s}}_{l}$ and the first principal component of the PCA decomposition in section \ref{pca}. For medium signal to noise (SNR) observations, the first principal component (PC), ie. the one with the highest eigenvalue associated to it, will contain the predominant lightcurve shape. As previously discussed, the first PC is not perfectly separated from the systematic signals and hence cannot be used directly for further analysis but it is good enough to use it as lightcurve identification. The identified lightcurve signal is labeled $\bf{\hat{S}}_{a}$. 

2) Once the lightcurve signal is identified, we exclude this row from $\bf{\hat{S}}$ and proceed to classify the remaining signals with respect to their non-Gaussianity (ie. systematic noise sources). Here we use the Ljung-Box portmanteau test \citep[see Appendix and][]{brockwell06} to test for the hypothesis that the time series is statistically white (ie. Gaussian). This test was originally designed to check the residuals of auto-regressive moving-average (ARMA) models for significant departures from Gaussianity.  It is hence ideally suited for our need to identify which estimated signal components are the desired non-Gaussian ones.   

The identified non-Gaussian, systematic noise, signals are hence labeled $\bf{\hat{S}}_{sn}$ and the remaining white noise signals $\bf{\hat{S}}_{wn}$ to give 

\begin{equation}
\bf{\hat{S}}_{a}  + \bf{\hat{S}}_{sn} + \bf{\hat{S}}_{wn} \overset{\triangle}{=}   \bf{\hat{S}} = \bf{\tilde{W}}\bf{\tilde{X}} 
\label{noisemodel}
\end{equation}

\noindent where $\bf{\hat{S}}$ has dimensions $k \times l$ and  $\bf{\hat{S}}_{a}$, $\bf{\hat{S}}_{sn}$, $\bf{\hat{S}}_{wn}$, have dimensions of $k \times 1$, $k \times l_{sn}$ and $k \times l_{wn}$ respectively where $ l = \sum l_{sn} + \sum l_{wn} + 1$.  The de-mixing matrix is given by $\bf{\tilde{W}} = \bf{\tilde{A}}^{-1}$. 

As previously mentioned, the components of $\bf{\hat{S}}$ have ambiguities in scaling and sign and can be thought to be similar to the eigenvectors of a principal component analysis with missing eigenvalues. Fortunately there exist two approaches to resolving this degeneracy: 

\begin{enumerate}
\item In the case of $\bf{\hat{S}}_{a}$ being well separated as individual component, we can take $\bf{\hat{S}}_{a}$ and the de-mixing matrix $\bf{\tilde{W}}$ and only retain the row containing the astrophysical signal component forming the row-vector $\bf{\tilde{w}_{a}}$. We then reconstruct the original data $\bf{\tilde{X}}$ using only the separated signal component: 

\begin{equation}
\bf{\tilde{X}_{a}} = \bf{\tilde{w}_{a}}^{-1} \bf{\hat{S}}_{a} = \bf{\tilde{w}_{a}}^{-1}  \bf{\tilde{W}}\bf{\tilde{X}} 
\label{sigcomp}
\end{equation}

\noindent where $\bf{\tilde{X}_{a}} $ is the reconstructed whitened data with all but the astrophysical signal components removed. Using equation \ref{whitening}, we can now calculate the un-whitened matrix $\bf{X_{a}}$.



\begin{align}
\label{sigcomp2}
\bf{X}_{a} &= \bf{Z}(\bf{X} - \bf{\bar{X}}) + \bf{\bar{X}} \\
\bf{Z} &= \bf{\tilde{w}}_{a}^{-1}\bf{\tilde{W}} 
\end{align}

\noindent Hence we can think of $\bf{Z}$ as a linear, optimal filter for the signal component in $\bf{X}$. Please note that this linear filtering does not impair the scaling information as this is re-instated going from $\bf{\hat{S}}_{a}$ to $\bf{{X}_{a}}$.
  
\item In the case of  $\bf{\hat{S}}_{a}$ not being well separated but other systematic noise components are, a different, more indirect approach can be used. Here, the systematic noise components, $\bf{\hat{S}}_{sn}$ which do not contain sign or scaling information, are simultaneously fitted to the time series data (preferably out-of-transit data), $x_{k}$. We therefore define the systematic noise model for an individual time series by $\bf{M}_{sn}$, 

\begin{equation}
\bf{M}_{sn} = \bf{O} \bf{\hat{S}}_{sn}
\label{noisemodel2}
\end{equation}

\noindent where $\bf{O}$ is a k $\times$ k diagonal scaling matrix of $\bf{\hat{S}}_{sn}$, which needs to be fitted iteratively as free parameters in the following section. 
\end{enumerate}

\subsection{Lightcurve fitting}
\label{fitting}

Having either filtered the data to obtain $\bf{X_{a}}$ or constructed the noise model $\bf{M}_{sn}$, we can now fit the original time series, $x_{k}$ using the standard analytical lightcurve models \citep{mandel02,seager03} in addition to the diagonal matrix $\bf{O}$, if necessary. For the purpose of this paper, which focuses on blind-source-separation, we will restrict ourselves to demonstrating the feasibility of estimating $\bf{O}$ and only leave the transit depth as variable lightcurve parameter. We use the analytic lightcurve model by \citet{mandel02} and a Nelder-Mead minimisation algorithm \citep{press07}. For real data applications, we advise the reader to use Markov Chain Monte Carlo methods, or similar, which have become standard in the field of exoplanets and allow the estimation of the posterior probability distributions and their associated errors \citep{bakos07, burke07, cameron07, ford06,gregory11}.

\subsection{Post-analysis}

Once the model fitting stage has been completed, we are left with fitting residual, $r_{k}$, i.e. $r_{k} = x_{k} - m_{k}$. Several tests are useful to determine how well the signals have been removed from the original time series, $x_{k}$. In the case of a full Markov Chain Monte Carlo fitting, the posterior distributions of the fitting parameters may be used to asses the impact of the remaining systematic noise in the data when compared to a simulated data set only containing white noise. Portmanteau tests on individual time series are useful to test for non-white noise signals and allow a measure of the overall performance of the algorithm \citep{brockwell06}. Additionally, we can determine the Kullback-Leibler divergence \citep{kullback51} of our residual's probability distribution function (pdf) to an idealised Gaussian case. 

For the simulations and real-data examples presented in the following section, we have merely plotted the autocorrleation functions (ACF) of the residuals obtained to determine whether for a given lag, these are within the 3$\sigma$ confidence limit of the residual being dominated by white-noise \citep{brockwell06, davison09}. Here the ACF is given by:

\begin{align}
ACF(k,\tau) &= \frac{1}{m} \sum^{m-\tau}_{t=1} (r_{k,t} - \bar{r}_{k})(r_{k,t + \tau} - \bar{r}_{k}) \\\nonumber  
& \tau = 0,1,2,3,... m/2
\label{ACF}
\end{align}

\noindent where $m$ is the number of data points in the time series, $\tau$ the specific lag and the confidence intervals are given by $\pm \sigma/ \sqrt{m}$.

\section{Simulations}
\label{simu}

\begin{figure}
\epsscale{1.0}
\plotone{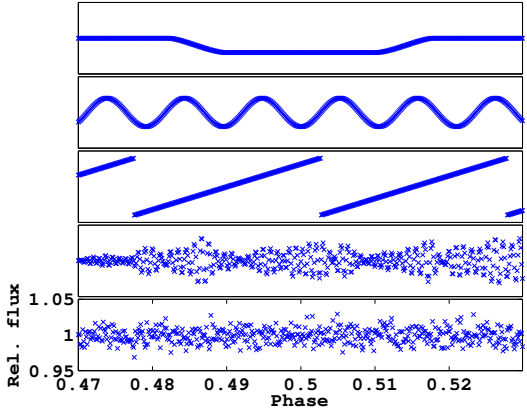}
\caption{Simulated input signals before mixing. From top to bottom: 1) secondary eclipse \citet{mandel02} curve, 2) sinusoidal function, 3) sawtooth function, 4) time-correlated auto-regressive function, 5) Gaussian noise. The scaling of the ordinate is identical for all subplots.  \label{ex1premix}}
\end{figure}

\begin{figure}
\epsscale{1.0}
\plotone{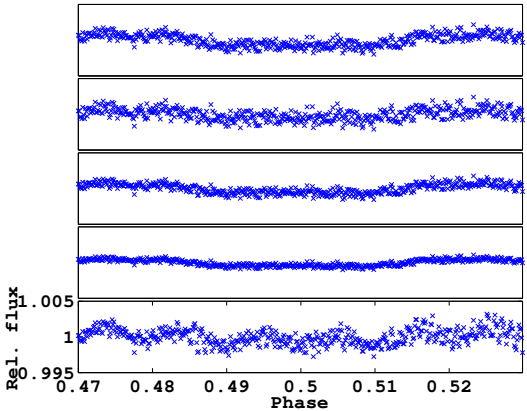}
\caption{The signals, $\bf{S}$, in figure \ref{ex1premix} were mixed using a random mixing matrix $\bf{A}$ to obtain the 'observed signals', $\bf{X}$ normalised to unity, shown in this diagram. The algorithm takes the lightcurves in this diagram as starting values. No further input is provided or assumptions on the underlying signals made. The scaling of the ordinate is identical for all subplots.  \label{ex1mix}}
\end{figure}

In order to test the behaviour and efficiency of the algorithm described above, we produced a toy model simulation with five observed signals: 1) a secondary eclipse \citet{mandel02} lightcurve; 2) a sinusoidal signal; 3) a sawtooth function; 4) a fourth order auto-regressive signal to simulate time-correlated signals; 5) Gaussian noise with a full width half maximum (FWHM) of 0.01 magnitudes. The premixed signals are displayed in figure \ref{ex1premix}. This gives us our signal matrix, $\bf{S}$, which needs  to be recovered later on. We have then proceeded to mix the signals in figure \ref{ex1premix} using a random mixing matrix, $\bf{A}$, to obtain our 'observed signals', $\bf{X}$, in figure \ref{ex1mix}. For the sake of comparability we keep the mixing matrix $\bf{A}$ to be the same for all simulations.

We now subdivide the simulations to illustrate the two possible methods of the signal reconstruction. {\it Method}~1 computes $\bf{X}_{a}$ using equation \ref{sigcomp2} whilst {\it Method}~2 fits the noise model $\bf{M}_{sn}$ (equation \ref{noisemodel2}) simultaneously with the \citet{mandel02} lightcurve. These two examples demonstrate that both techniques work equally well for a well behaved data set.

\subsection{Method~1: Filtering out the signal}

\begin{figure}
\epsscale{1.0}
\plotone{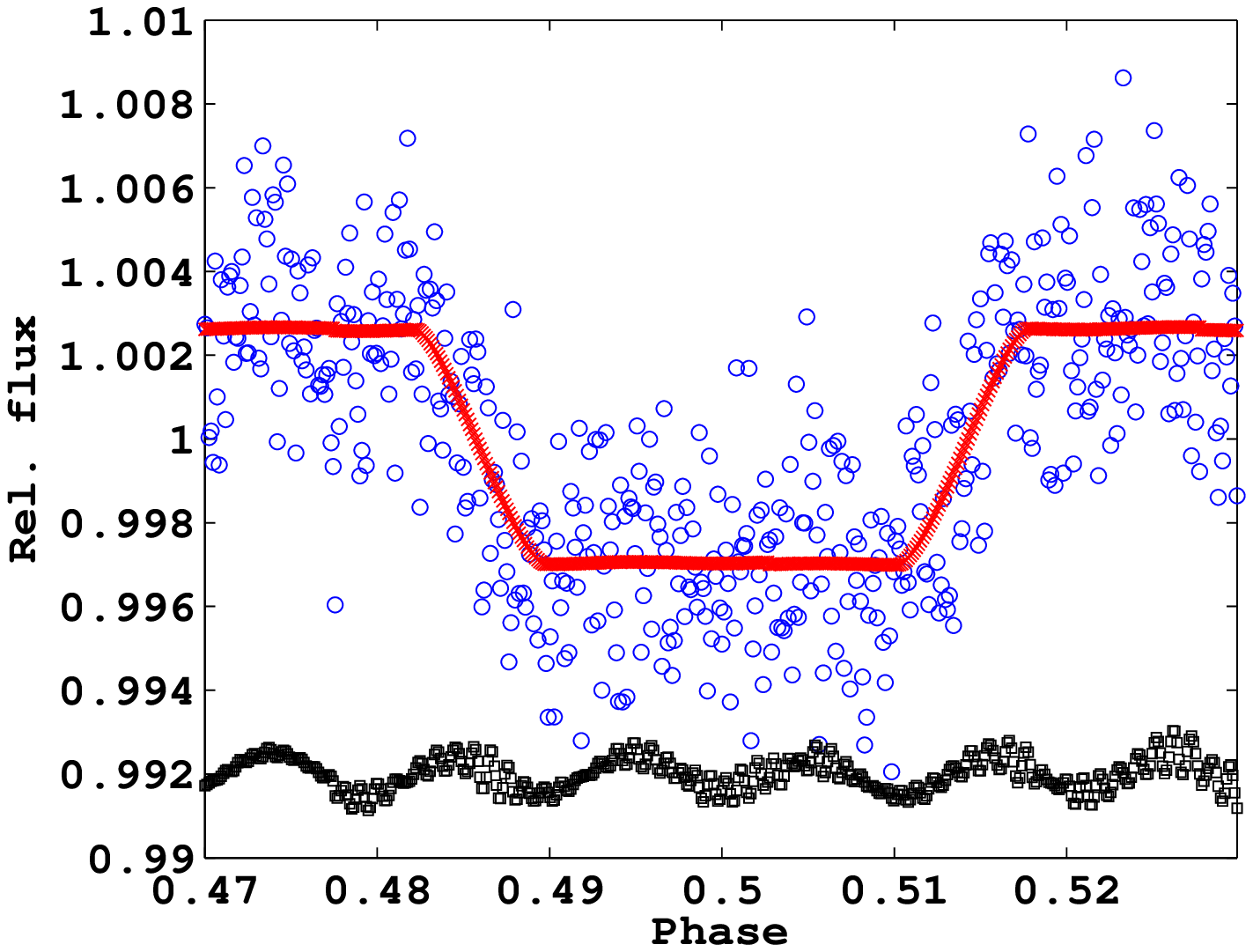}
\caption{Results of the blind-source separation. The blue circles present the the first lightcurve of the raw data $\bf{X}$, the red crosses the retrieved signal component, $\bf{X}_{a}$, and the black squares the systematic noise component $\bf{X}_{sn}$.  \label{ex1filter}}
\end{figure}

\begin{figure}
\epsscale{1.0}
\plotone{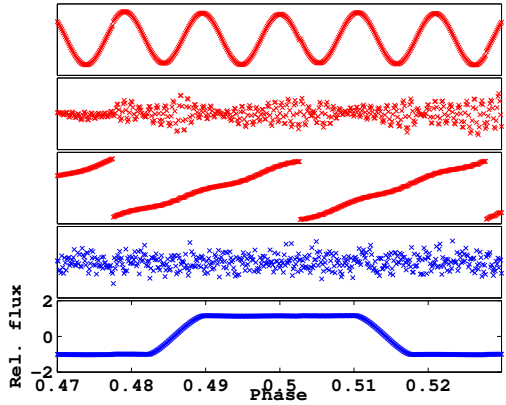}
\caption{Results of the blind-source separation. The top three signals in red were identified by the algorithm to comprise the systematic noise model, $\bf{\hat{S}_{sn}}$. The 4th signal was correctly identified to be Gaussian noise and the bottom to be the lightcurve signal. Note that the blind-source-separation does not preserve signs nor scaling of the estimated signals. The scaling of the ordinate is identical for all subplots. \label{ex1postmix}}
\end{figure}

In this example, we use the 'observed' signals in figure \ref{ex1mix} as input to the algorithm. We however do not perform a dimensionality reduction using PCA since we are not dealing with an over-complete set in this example. The results of the separation are shown in figure \ref{ex1postmix}. Here the top three, red lightcurves are the estimated systematic noise components as identified by the algorithm. The fourth component is Gaussian noise and the bottom is an inverse of the lightcurve signal. It should again be noted here that the blind-source separation does not preserve the scaling nor the signs of the signals in $\bf{\hat{S}}$. However, when the original data is reconstructed using only the signal component, $\bf{\hat{S}}_{a}$, to obtain $\bf{X}_{a}$ (equation \ref{sigcomp2}), the scaling and sign informations are re-instated. For a well behaved data set, i.e. one that obeys the instantaneous mixing model and has negligible Gaussian noise in their signal components, it is therefore possible to re-construct the lightcurve signal from the raw data as explained in section \ref{nmodel}. Figure \ref{ex1filter} shows the top lightcurve of figure \ref{ex1mix} (blue circles) and overplotted the retrieved signal component (red crosses) and offset below the systematic noise component (black squares).  

\begin{figure}
\epsscale{1.0}
\plotone{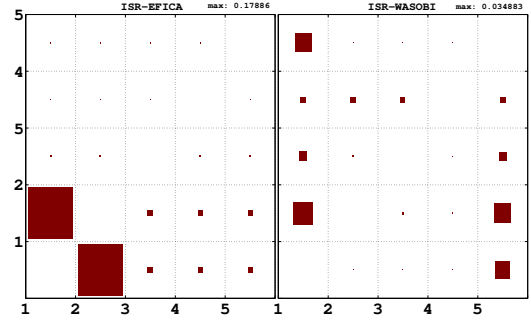}
\caption{Hinton diagram of the EFICA and WASOBI interference-over-signal matrices for Example~1. The polygon areas are normalised to the highest value in the matrix (given in the bottom corners). The smaller the off-diagonal elements of the matrix, the higher the signal separation efficiency of the algorithm. In this case we can see the EFICA algorithm to perform better than the WASOBI one. \label{ex1isrs}}
\end{figure}

As a useful by-product of the algorithm, we obtain the interference over signal matrices (ISR, equations \ref{ISRef} $\&$ \ref{ISRwa} in the Appendix \ref{bss}) for both the EFICA and WASOBI algorithms. These give us valuable information on the efficiency at which the signals have been separated. Figure \ref{ex1isrs} shows the Hinton diagrams of the EFICA and WASOBI $\bf{ISR}$ matrices. Here, the smaller the off-diagonal elements of the matrix, the better the signal separation. In this example, the EFICA algorithm outperforms the WASOBI one, which is to be expected since all signals but one are non-Gaussian. 

\subsection{Method~2: Fitting a noise model to the data}

\begin{figure}
\epsscale{1.0}
\plotone{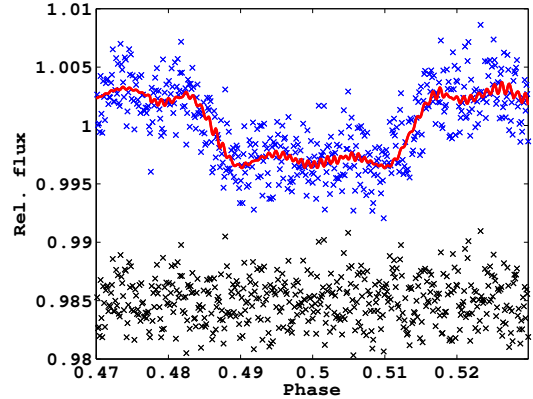}
\caption{showing the raw lightcurve (first row in figure \ref{ex1mix}, blue) normalised to unity, with the model fit (red) overlaid and the fitting residuals plotted underneath (black).   \label{ex1fitted}}
\end{figure}

\begin{figure}
\epsscale{1.0}
\plotone{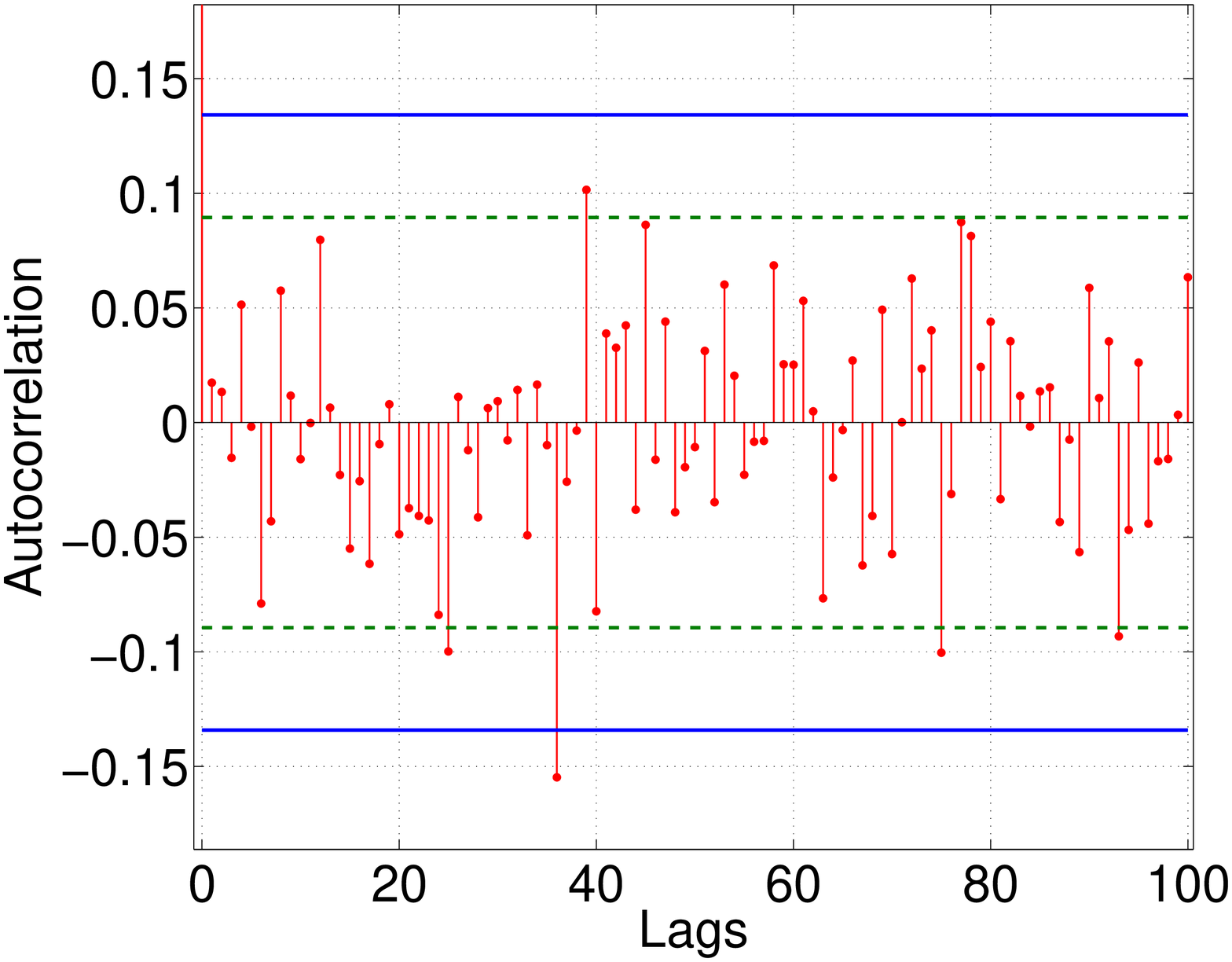}
\caption{showing the auto-correlation function for 250 lags (red). The 3$\sigma$ confidence limits that the observed residual is normally distributed are shown in blue. All but two lags are within the confidence limits, strongly suggesting that the residual is dominated by white noise and correlations were efficiently removed.     \label{ex1auto}}
\end{figure}

In the previous section, we have shown that in the case that the astrophysical component $\bf{\hat{S}}_{a}$ is well separated as individual signal, we can create a filter for the raw data that directly filters the lightcurve signal from the noise. However, in most real data applications, $\bf{\hat{S}}_{a}$, is not perfectly separated but the components of $\bf{\hat{S}}_{sn}$ may be more so. In this case we can construct the noise model $\bf{M}_{sn}$ given by equation \ref{noisemodel2}  and the diagonal elements of $\bf{O}$ are fitted as described in section \ref{fitting}. The starting position of the algorithm is the same as for the previous example (figure \ref{ex1mix}). The model fit of the first lightcurve in figure \ref{ex1mix} and its residuals are shown in figure \ref{ex1fitted}. The autocorrelation function for 100 lags is plotted in figure \ref{ex1auto}. All but two lags are within the 3$\sigma$ confidence limit that the residual is white noise dominated, indicating that all signals have been removed effectively. 
  
\begin{figure}
\epsscale{1.0}
\plotone{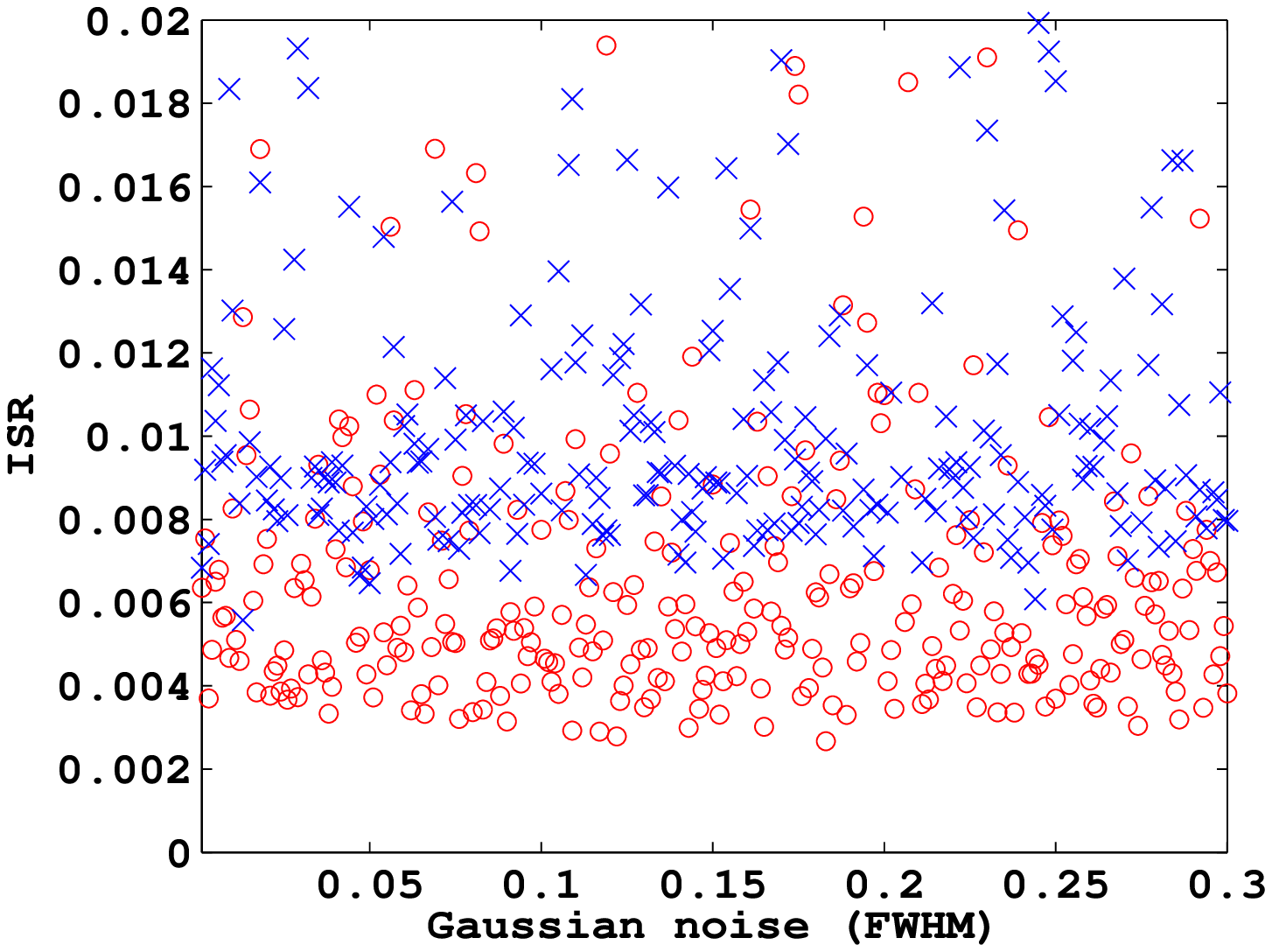}
\caption{showing the mean interference over signal ratios (ISRs) for both the EFICA (red circles) and WASOBI (blue crosses) algorithms for Example~1. In this example, the EFICA algorithm clearly outperforms WASOBI by reaching lower ISR values. Both algorithms are stable even under low signal to noise conditions.\label{ex1ISR2}}
\end{figure}

Finally we simulate the convergence properties of both EFICA and WASOBI under varying white noise conditions. Here we repeatedly run the algorithm until signal separation is completed and record the mean ISRs of the source separation. We performed this simulation 300 times for Gaussian noise FWHMs varying from 0.0 - 0.3 magnitudes (figure \ref{ex1fitted} has a FWHM$_{gauss}$ = 0.01) and every ISR measurement reported is the mean of 10 iterations.  Figure \ref{ex1ISR2} summarises the results. Here, the red circles represent the mean ISR or the EFICA algorithm and the blue crosses that of WASOBI. It can clearly be seen that for this example the EFICA algorithm outperforms WASOBI and on average reaches lower ISR values. We can further note that the blind source separation is not significantly affected by the magnitude of the white noise and performs well under difficult signal to noise conditions. 
  
\section{Application to data}

The previous examples illustrated the two methods available to correct the observed data: {\it Method 1}: Filtering the astrophysical signal out of the systematic noise or {\it Method 2}: fitting a systematic noise model to the raw data.  Here we apply these techniques to two primary eclipse data sets of HD~189733b and XO1b recorded by the NICMOS instrument on the Hubble Space Telescope as well as a single time-correlated time series obtained by the Kepler spacecraft .  

\subsection{HST/NICMOS: HD~189733b}
\label{hd189sect}

\begin{figure}
\epsscale{1.0}
\plotone{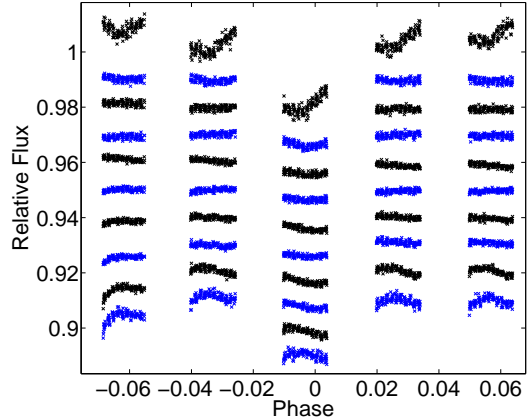}
\caption{showing 'raw', extracted HST/NICMOS light-curves of HD~189733b primary eclipse. Light curves are offset for clarity. \label{hd189rawlc}}
\end{figure}

First presented by \citet{swain08}, this data set of the primary eclipse of HD~189733b was recorded using HST/NICMOS in the G206 grism setting spanning five consecutive orbits. The HST-pipeline calibrated data were downloaded from the MAST\footnote{http://archive.stsci.edu/} archive and the spectrum was extracted using both standard IRAF\footnote{http://iraf.noao.edu/} routines as well as a custom built routine for optimal spectral extraction. Both extractions are in good accord with each other but the custom built routine was found to yield a better signal to noise and was subsequently used for all further analysis. A binning of 10 spectral channels ($\sim$ 0.08$\mu$m) was used resulting in 10 light curves across the G206 grism band. Figure \ref{hd189rawlc} shows the obtained time series which serve as input to the algorithm. It can be seen that each time series is strongly affected by instrument systematics propagating from the blue side of the spectrum (bottom light curve) to the red with varying intensity and even sign. \citet{swain08} showed that these systematics are correlated to instrument state vectors such as orbital phase, relative positions and angles of the spectrum on the detector, instrument temperature, etc. We can hence expect that these systematics are statistically independent from the recorded astrophysical signal (the light curve) and it should therefore be in principle possible to de-correlate the signal. 

\begin{figure}
\epsscale{1.0}
\plotone{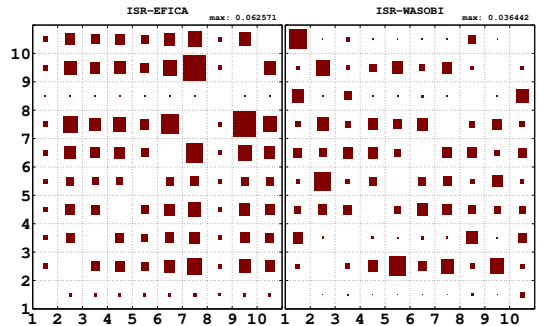}
\caption{the Interference over Signal (ISR) matrix of the component separation for both the EFICA and the WASOBI algorithms. All values were normalised with the maximum ISR = 0.0626. Components 1, 3, 5 $\&$ 8 yielding the lowest ISR values and correspond to the astrophysical light curve signal (comp. 1) and the three most prominent systematic noise vectors in figure \ref{hd189snvec}. Other components were identified as predominantly Gaussian or weakly systematic by the pipeline.  \label{hd189hinton}}
\end{figure}

We here demonstrate the de-trending on an individual light curve at $\sim$1.694$\mu$m (8$^{th}$ one down in figure \ref{hd189raw}). All time series in figure \ref{hd189raw} were taken as input to the algorithm described above to estimate the de-mixing matrix $\bf{\tilde{W}}$, the astrophysical signal vectors, $\bf{\hat{S}_{a}}$ and the systematic noise vectors, $\bf{\hat{S}_{sn}}$.  The interference over signal (ISR) matrix indicated the good separation of four main components figure \ref{hd189hinton} with the rest of the components being classified as predominantly Gaussian or weakly systematic. The existence of more than one Gaussian component ($l_{wn} > 1$) indicates that the set is overcomplete. However since the data-set is small enough, no PCA dimensionality reduction was performed. After the algorithm has identified the correct astrophysical signal, it proceeded to reconstruct the light curve using both methods described above. 

\begin{figure}
\epsscale{1.0}
\plotone{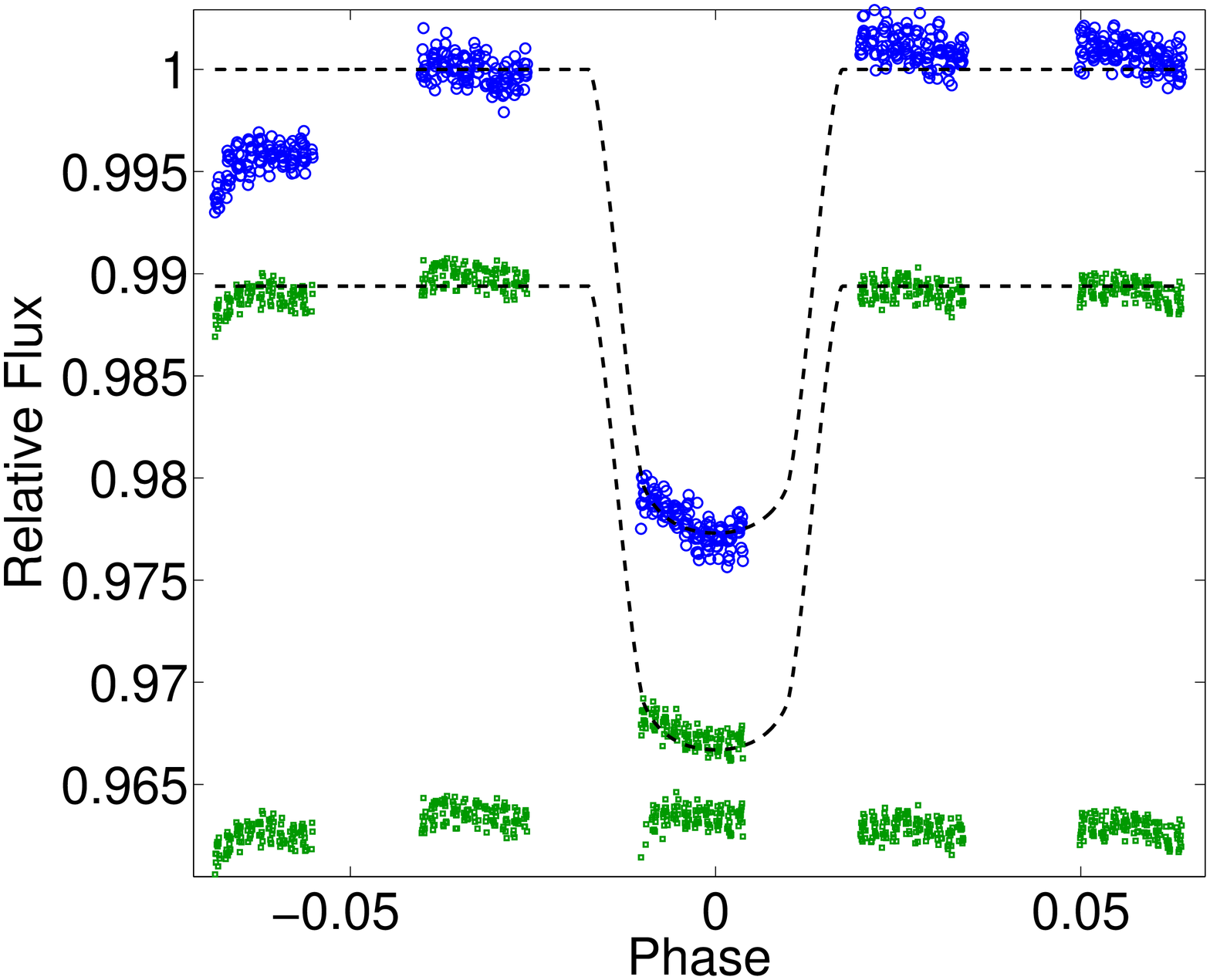}
\caption{showing the raw-data light curve (blue crosses) and the corrected light curve (green squares) offset below.  In this example, we used equations \ref{sigcomp} $\&$ \ref{sigcomp2} as light curve filter. The systematic noise components were reduced but residual systematics remain in the final light curve.  \label{hd189componly}}
\end{figure}

\begin{figure}
\epsscale{1.0}
\plotone{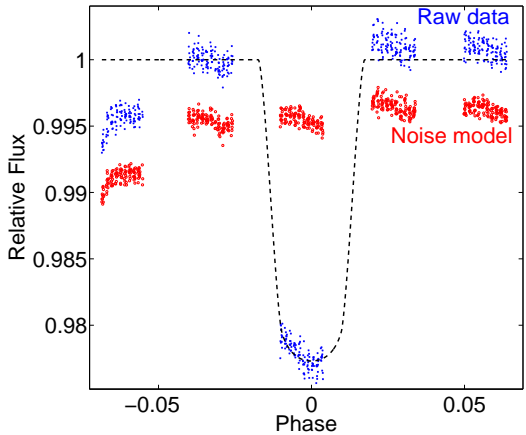}
\caption{showing the same 'raw' light curve as in Figure \ref{hd189componly} (blue crosses) and the calculated systematic noise model (red circles) offset below.  \label{hd189raw}}
\end{figure}

{\it Method 1}: The astrophysical signal was filtered using equations \ref{sigcomp} $\&$  \ref{sigcomp2}. Figure \ref{hd189componly} shows the raw light curve (blue circles) with the de-trended time series, $\bf{X}_{a}$ underneath (green squares). Superimposed light curves were computed using \citet{mandel02} with orbital parameters were taken from \citet{winn07} and limb-darkening parameters from \citet{claret00}. It is clear that the de-trended light curve is an improvement to the raw time series but that systematics still remain in the data. This is further illustrated by plotting the autocorrelation function of the model-fit residual in figure \ref{hd189acf} (red circles). Here, residual correlation can be observed in particular at low lags. This is a consequence of the astrophysical signal, $\bf{\hat{S}_{a}}$, being well separated but as shown in figure \ref{hd189hinton} (component 1), there remains some weak interference between the  $\bf{\hat{S}_{a}}$ and other vectors, which is a consequence of equation \ref{demix2} and to be expected for real data-sets.

\begin{figure}
\epsscale{1.0}
\plotone{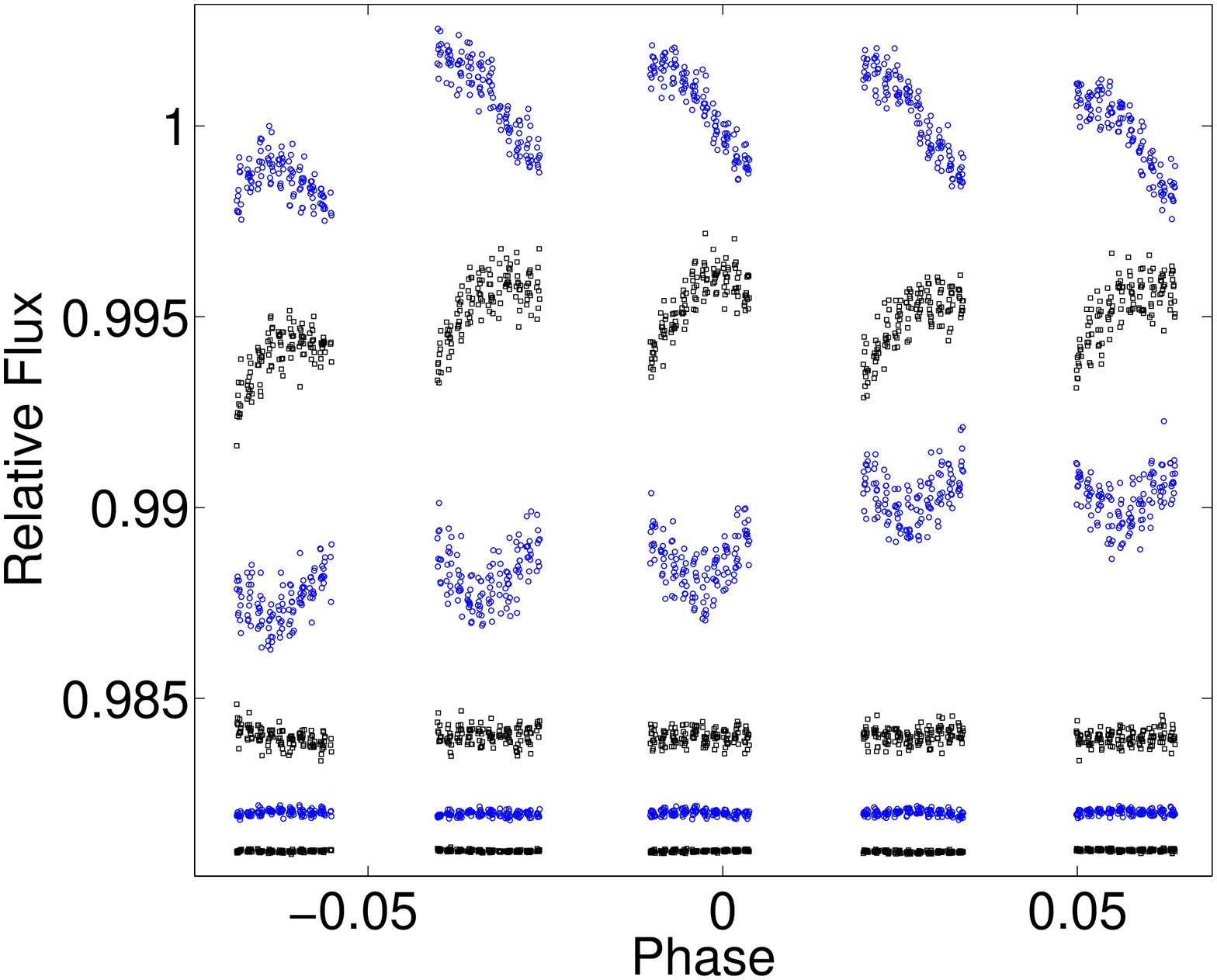}
\caption{Individual systematic noise vectors, $\bf{\hat{S}_{sn}}$, of HD~189733b, with the appropriate scaling. Combined they form the systematic noise model in figure \ref{hd189raw} (red circles).  \label{hd189snvec}}
\end{figure}

\begin{figure}
\epsscale{1.0}
\plotone{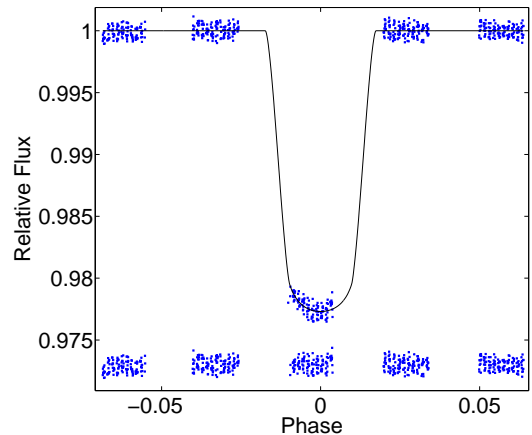}
\caption{showing the de-trended data by subtracting the noise model of the raw data.  \label{hd189corr}}
\end{figure}

\begin{figure}
\epsscale{1.0}
\plotone{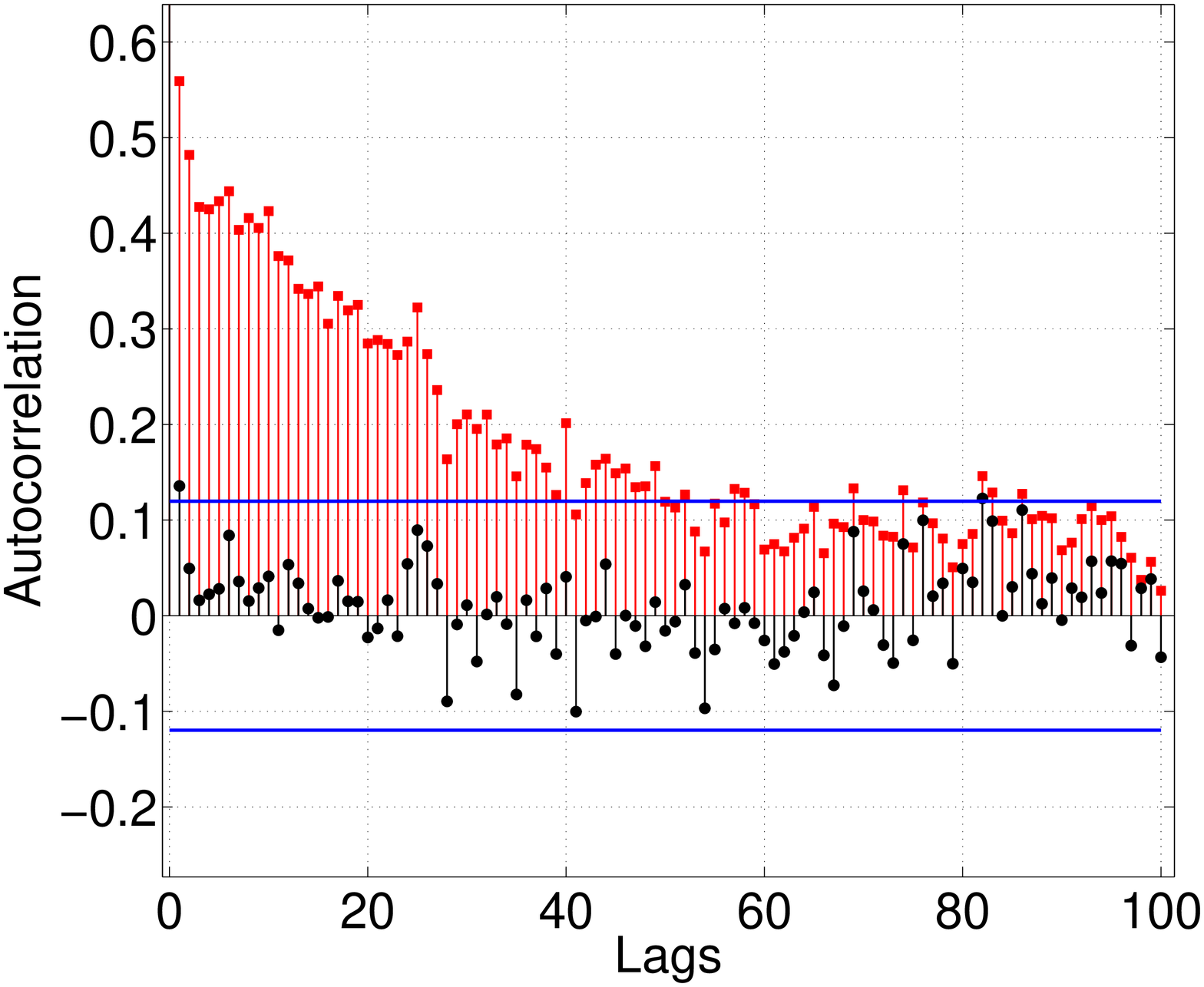}
\caption{showing the autocorrelation function for 100 lags of the fitting residual in figure \ref{hd189componly} (red squares) and figure \ref{hd189corr} (black circles). The blue lines signify 3$\sigma$ limits for a Gaussian distribution. The fitting residual of figure \ref{hd189componly} shows high amounts of residual correlation, particularly at lower lags whilst the fitting residual of figure \ref{hd189corr} follows a Gaussian distribution.  \label{hd189acf}}
\end{figure}

{\it Method 2}: The second method is a less direct approach. Instead of filtering for the astrophysical signal directly, we try to construct a 'systematic noise model' that is then subtracted off the raw data. Using equation \ref{noisemodel2} and a simplex downhill algorithm \citep{nelder65} we estimated the scaling matrix, $\bf{O}$, by fitting the the systematic noise vectors, $\bf{\hat{S}_{sn}}$ to the four out of transit orbits. The scaled systematic noise vectors are shown in figure \ref{hd189snvec} which combine to form the systematic noise model, $\bf{M}_{sn}$, in figure \ref{hd189raw}. It should be noted that $\bf{O}$ is only a scaling matrix of the individual vectors as the scaling information is not preserved by the independent component analysis. Hence, relative intra and inter-orbit variations are preserved. Figure \ref{hd189corr} shows the corrected data by subtracting the systematic noise model off the raw data. Inspecting the fitting residual's autocorrelation function in figure \ref{hd189acf} (black circles) indicates the residual to be statistically white and a maximal de-correlation of the data has been achieved.

\subsection{HST/NICMOS: XO1-b}

\begin{figure}
\epsscale{1.0}
\plotone{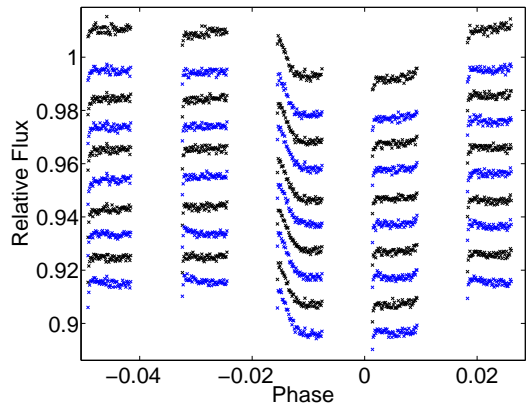}
\caption{showing 'raw', extracted HST/NICMOS light-curves of HD~189733b primary eclipse. Light curves are offset for clarity. \label{xo1rawlc}}
\end{figure}

\begin{figure}
\epsscale{1.0}
\plotone{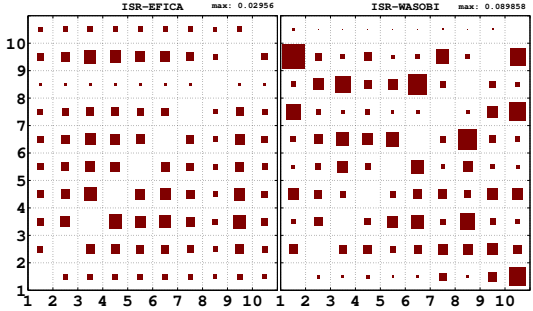}
\caption{same than for figure \ref{hd189hinton}. The light curve vector (component 1) shows residual interference with other vectors for both EFICA and WASOBI algorithms. Overall the EFICA algorithm outperforms WASOBI.  \label{xo1hinton}}
\end{figure}

Originally presented by \citet{tinetti10}, the primary eclipse of XO1b was observed using the HST/NICMOS instrument in the G141 grism setting.  The HST-pipeline calibrated data was downloaded and the spectra extracted using the same settings as for section \ref{hd189sect}. This yielded 10 light curves and which serve as input to the algorithm, see figure \ref{xo1rawlc}. Similar to the HD~189733b the algorithm retrieved four main components, the light curve signal and three main systematic noise components. The ISR matrix can is shown in figure \ref{xo1hinton}. We now proceeded to de-trending the light curve at the very red end of the spectrum (first from top in figure \ref{xo1rawlc}) as it, after visual inspection, exhibits the most prominent systematics of the 10 time series. Light curve fits assumed limb-darkening and orbital parameters by \citet{burke10}. 

\begin{figure}
\epsscale{1.0}
\plotone{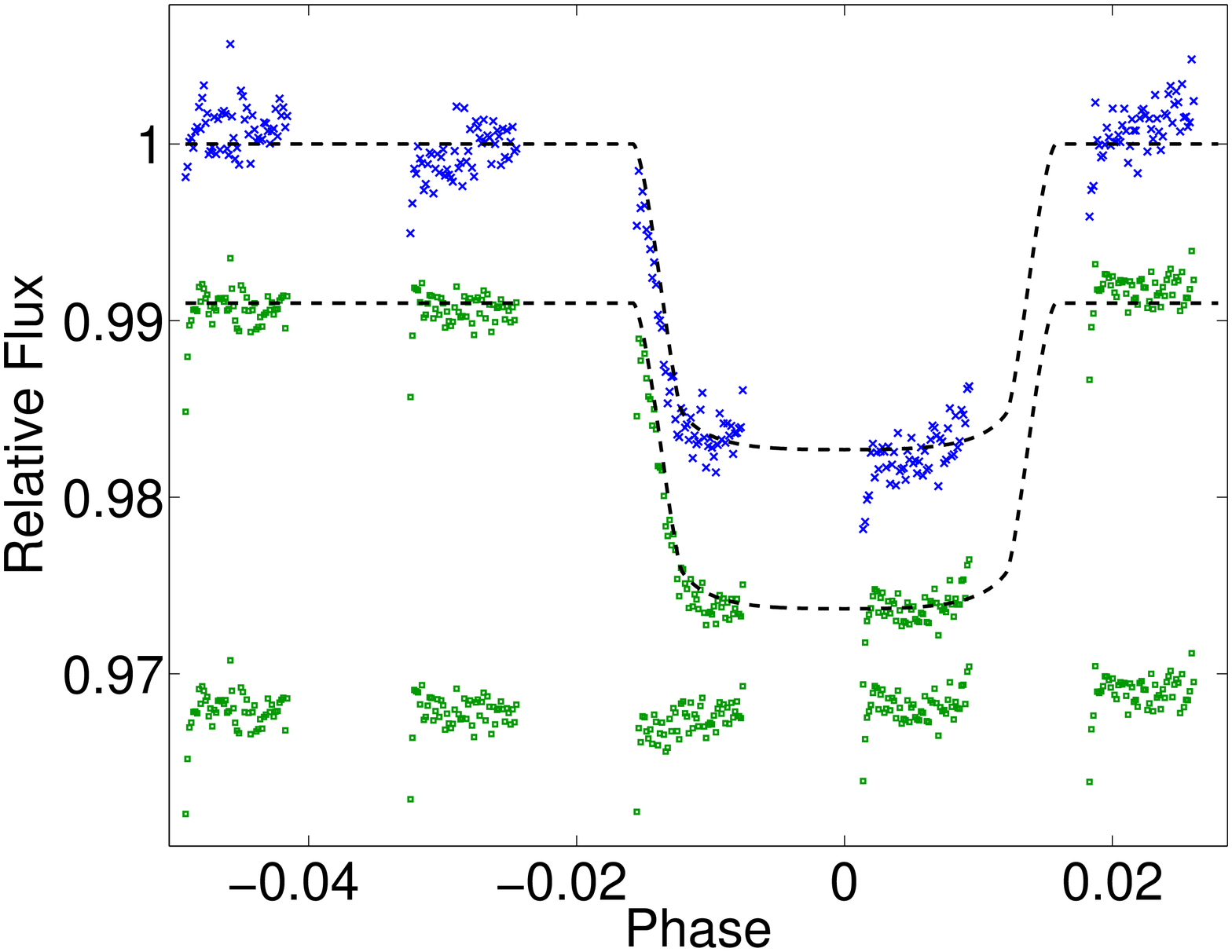}
\caption{showing the raw-data light curve (blue crosses) and the corrected light curve (green squares) offset below.  In this example, we used equations \ref{sigcomp} $\&$ \ref{sigcomp2} as light curve filter. The systematic noise components were reduced but residual systematics remain in the final light curve.  \label{xo1componly}}
\end{figure}

\begin{figure}
\epsscale{1.0}
\plotone{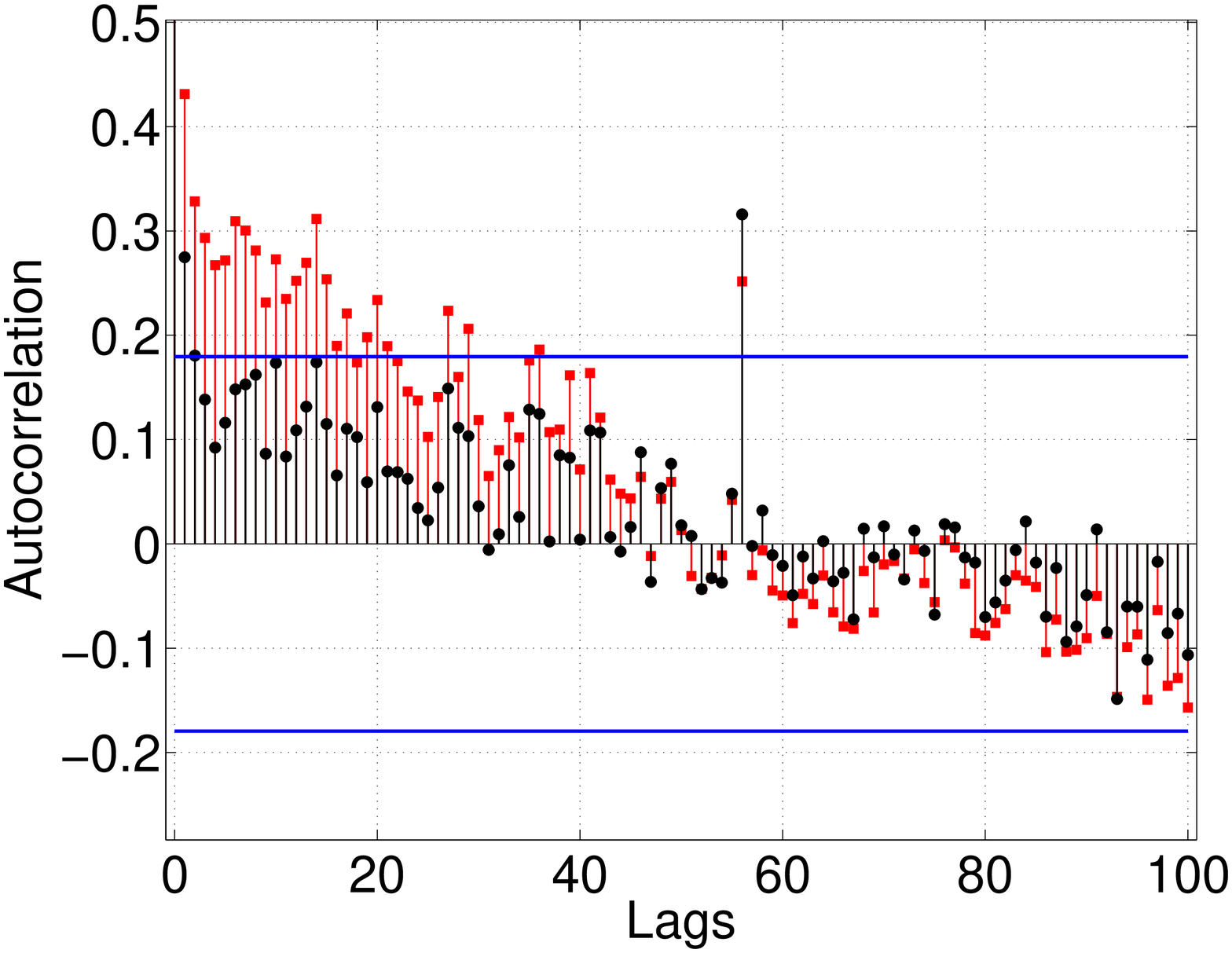}
\caption{showing the autocorrelation function for 100 lags of the fitting residual for method~1 (red squares) and method~2 (black circles). The blue lines signify 3$\sigma$ limits for a Gaussian distribution. The fitting residual of method~1 shows residual correlation, particularly at lower lags whilst the fitting residual of method~2 is by a factor of two better de-correlated in the lower lags.  \label{xo1acf}}
\end{figure}

{\it Method~1}: Figure \ref{xo1componly} shows the raw time series and the de-trended light curve using equation \ref{sigcomp2}. The light curve is significantly de-trended but systematics remain in the data as also shown by the autocorrelation function (red circles) in figure \ref{xo1acf}.  

\begin{figure}
\epsscale{1.0}
\plotone{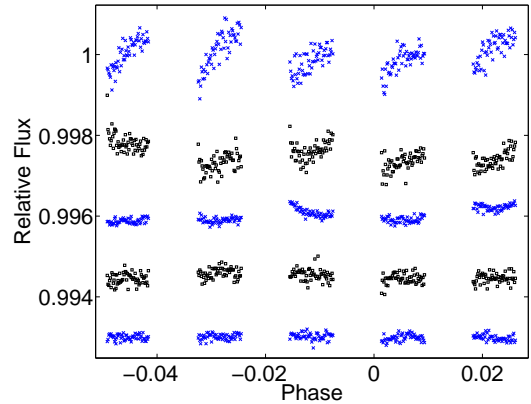}
\caption{Individual systematic noise vectors, $\bf{\hat{S}_{sn}}$, of XO1b, with the appropriate scaling. Combined they form the systematic noise model in figure \ref{hd189raw} (red circles).  \label{xo1snvec}}
\end{figure}

\begin{figure}
\epsscale{1.0}
\plotone{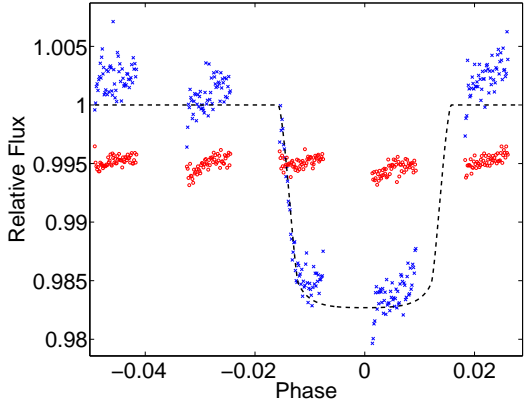}
\caption{showing the same 'raw' light curve as in Figure \ref{xo1componly} (blue crosses) and the calculated systematic noise model using the systematic noise vectors in figure \ref{xo1snvec}.  \label{xo1raw}}
\end{figure}

{\it Method~2}: As described in previous sections, figure \ref{xo1snvec} shows the retrieved systematic noise vectors and figure \ref{xo1raw} features the 'raw' data with the combined systematic noise model (red) underneath. The autocorrelation function of the model fit residual is shown in figure \ref{xo1acf} (black crosses) and shows a factor 2 improvement on the de-correlation in the lower lags. 

\begin{figure}
\epsscale{1.0}
\plotone{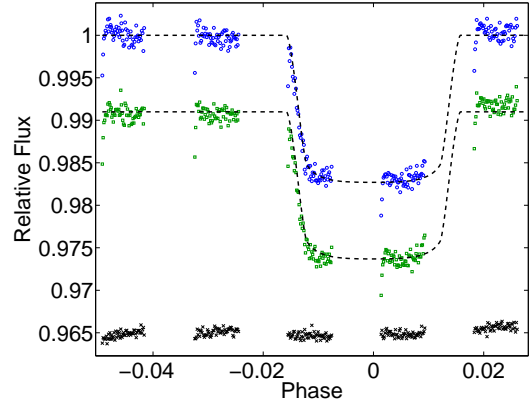}
\caption{showing the de-trended data using method~1 (top blue circles) and method~2 (bottom green squares) offset from each other. Both results show little differences between them as seen by the residual of method~1 - method~2 (black crosses).    \label{xo1comp2}}
\end{figure}

Figure \ref{xo1comp2} compares the de-trended light curves of {\it method~1} and {\it method~2} and shows the residual of {\it method~1} - {\it method~2} (black crosses). There is little difference between both methods indicating that the signal separation for this data-set is close to its maximum with the data being partially de-correlated. This is in contrast to the HD~189733b example where a near perfect de-correlation was achieved and can be attributed to the systematics being mostly wavelength invariant in the case of XO1b. In other words, systematic noise components which have a constant weighting throughout the data set cannot be de-correlated using ICA or PCA methods, which is to be expected following equations \ref{intro1} - \ref{timeseries3}.

\subsection{Kepler: Star 10118816}
\label{kepler}

In the previous examples we have shown that spectroscopic datasets can be de-correlated effectively. We test here how well the proposed algorithm can de-correlate consecutively observed data-sets. This is of particular interest in cases where no multi-channel data are available and the time series data are contaminated with time-correlated noise, be it stellar or instrumental. 
As opposed to the previous examples, where several time series, $x_{k}$, were observed simultaneously, here we take a single time series covering several consecutive eclipse features and cut the time series into segments spanning equal lengths over each eclipse event. Using these segments as inputs to the algorithm clearly violates the underlying assumptions of the independent component analysis, as the mixing is not instantaneous. In this case, the ICA analysis can be understood as a Projection Pursuit (PP) analysis, see section~\ref{projectionpursuit} and \citep{hyvarinen00, icabook3, huber85}. Here the ICA algorithm, in the absence of a working ICA data-model, will try to extract as many non-Gaussian components as possible and return the rest of the data in its original form. 
This is very similar to Projection Pursuit, where the data is not described by an underlying data model at all but only the most non-Gaussian component is retrieved. In other words, we can only expect to retrieve the eclipse signal component, $\bf{\hat{S}}_{a}$, with any degree of accuracy. As a result we will not be able to retrieve systematic noise components, $\bf{\hat{S}}_{sn}$, and we can only use $Method~1$ (in section \ref{nmodel}) to de-trend the data.

\begin{figure}
\epsscale{1.0}
\plotone{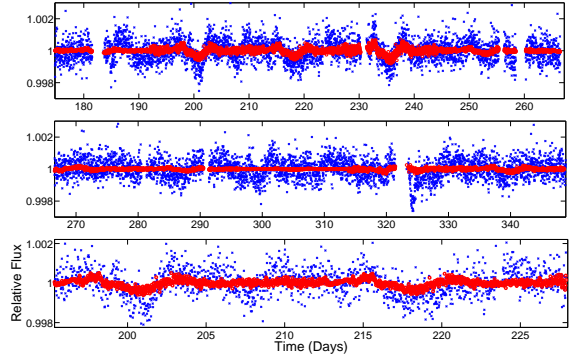}
\caption{Input time series (blue crosses) with filtered signal using $Method~1$ over plotted (red circles). Bottom plot is a zoomed in part of the time series above.The algorithm effectively filtered for the desired feature and strongly decreased contributions from autocorrelated noise.  \label{kepler3}}
\end{figure}

We have downloaded data observed by the Kepler space telescope \citep{borucki96, borucki10, jenkins10, caldwell10, koch10} for a planet-hosting candidate star observed over the second and third data-release quarters (Q2 \& Q3). The time series, with the Kepler ID: 10118816, exhibits highly variable features and significant time-correlated noise (see figure \ref{kepler3}, blue crosses). Given Kepler's superb instrument calibration, we can assume this time-correlated noise to be due to stellar variability. Using the periodogram calculator on the NSteD database\footnote{$http://nsted.ipac.caltech.edu/applications/ETSS/Kepler\_index.html$}, we identified four main periodically recurring signals in the data-set. Choosing the second strongest feature with a period of 0.040915 days, we phase folded the data and cut the time series in 10 equally sized segments. As for the previous examples we now took these time series segments as input to the algorithm. 

\begin{figure}
\epsscale{1.0}
\plotone{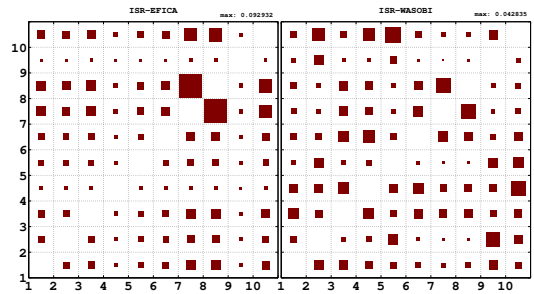}
\caption{the Interference over Signal (ISR) matrix of the component separation for both the EFICA and the WASOBI algorithms. All values were normalised with the maximum ISR = 0.09293. Components 4 and 9 are the best separated, with component 4 being the desired signal component.   \label{kepler4}}
\end{figure}

\begin{figure}
\epsscale{1.0}
\plotone{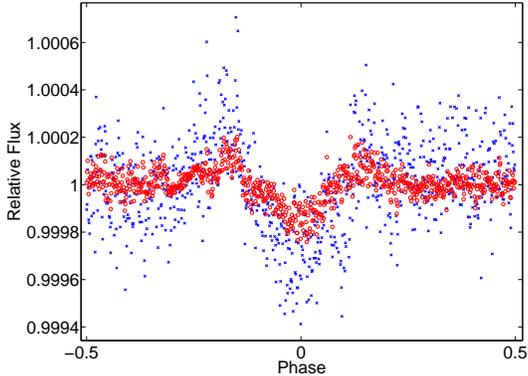}
\caption{showing the mean, phase-folded feature (blue crosses) with the ICA filtered signal component (red circles) over plotted. The ICA filtered signal shows a significant reduction in scatter and auto-correlative noise compared to the simply phase folded data. \label{kepler1}}
\end{figure}

\begin{figure}
\epsscale{1.0}
\plotone{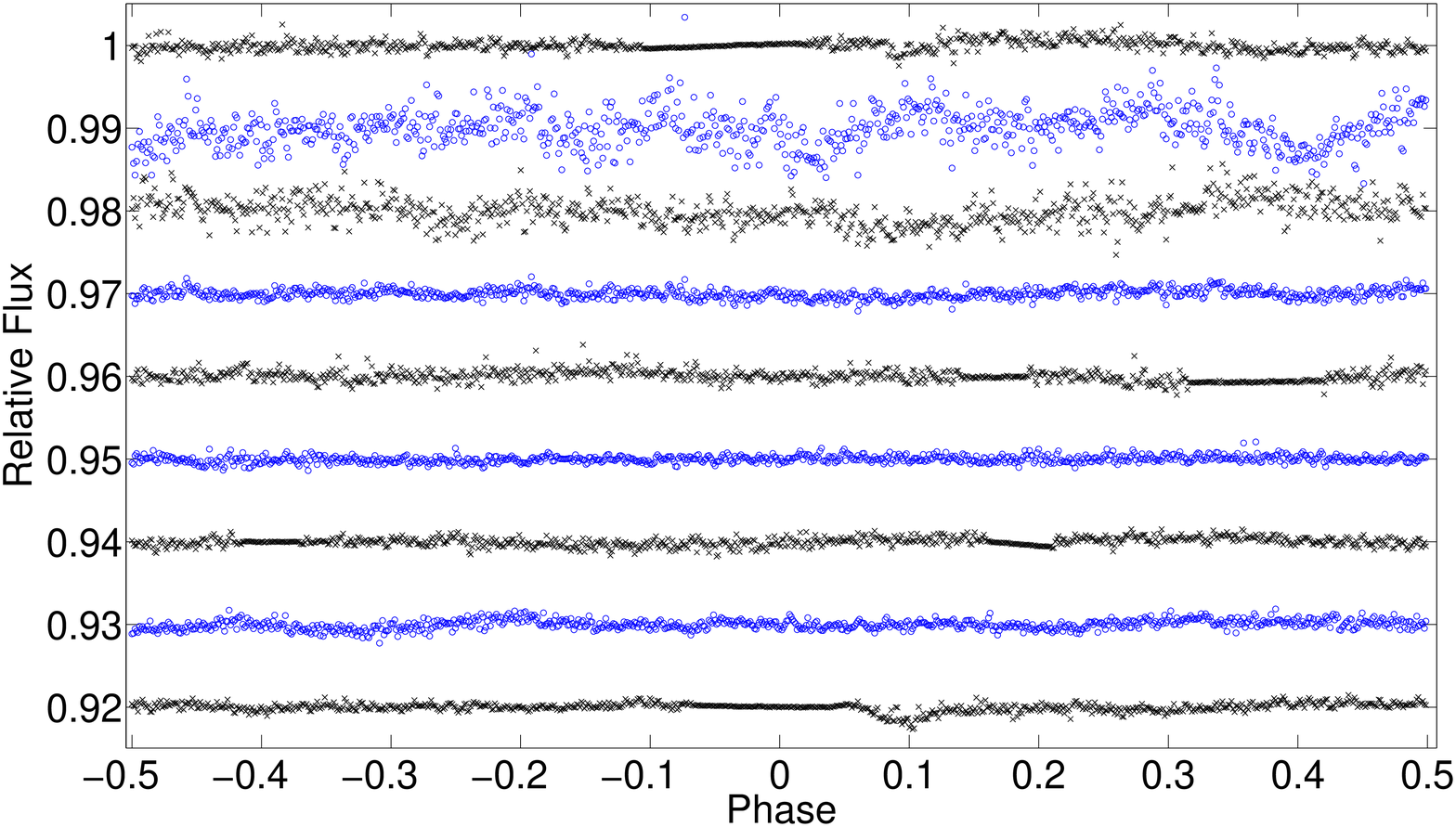}
\caption{Addition components to the signal in figure \ref{kepler1} as calculated by the algorithm.  \label{kepler2}}
\end{figure}

We performed our de-correlation as for the previous examples but using {\it Method~1} only. Figure~\ref{kepler4} shows the ISR matrix of the separation indicating a relatively poor separation of the components but two (components 4 \& 9). As discussed above, this behaviour is to be expected with the breaking of the instantaneous mixing model. Nonetheless, we obtained a clear feature (component 4) in our analysis which is over plotted (red circles) on the mean, phase-folded data (blue crosses) in figure~\ref{kepler1}. Here the de-correlated signal has a much reduced scatter compared to the mean of the phase-folded feature, which indicates that much of the unwanted stellar variability has been removed. It is also clear from this figure that we are not dealing with an exoplanetary light curve but a stellar pulsation feature. 
As expected, the remaining components returned from the algorithm (figure~\ref{kepler2}) are the residuals of the input data minus the component shown in figure~\ref{kepler1}. Hence we only used the component in figure~\ref{kepler1} to re-construct the original time series. This was done by using equation \ref{sigcomp2} on each segment of the time series, followed by adding the segments back together in the order they were originally split up.  

Figure~\ref{kepler3} shows the original input data (blue crosses) with the filtered signal (red circles) over plotted on top. In the bottom plot (a zoomed in version of the time series), it is clear that the desired feature remains in the filtered time series whilst the contribution of other time-correlated stellar noise is substantially reduced.

\section{Discussion}

In the previous sections we have shown that for a set of simultaneously observed time series data (e.g. following an exoplanetary eclipse with a spectrograph)  we can describe the data by an instantaneous mixing model (equation \ref{timeseries3}). This allows the separation of non-Gaussian, time and spatially-correlated signals from one another. The degeneracy caused by not being able to retrieve the component's signs or amplitudes can be circumvented in two ways: {\it Method 1}) The separated signals are used to construct a linear transformation to filter the astrophysical signal from the originally observed data and hence preserve all scaling information; {\it Method 2})  The separated astrophysical signal is not used directly but instead all systematic noise components are combined to form a `systematic noise model' which can then be used to correct the original observed data. 

We have explored the efficiency of the signal de-trending on two simulated and two HST/NICMOS data sets with different types of systematic noise due to different grisms. The simulations demonstrate the two methods of de-trending the data in an idealised case and explore the efficiency of the signal separation in the presence of varying Gaussian noise in the data. In the instantaneous mixing model employed here, Gaussian noise sources are only indirectly allowed and can interfere with the effectiveness of separating non-Gaussian vectors. We tested this point by adding additional Gaussian noise components of variable amplitude to the simulations but did not observe any significant reductions in the signal separation efficiency.  

We proceeded to analyse two HST/NICMOS data sets: the primary eclipses of HD189733b and XO1b. For both data sets we find the {\it Method 2} to yield better results. In the case of HD189733b, we can achieve a near perfect de-correlation of astrophysical signal and systematic noise and no further steps are necessary to the de-correlation process. A more in depth discussion of this data set and HST/NICMOS systematics is beyond the scope of this publication. In the case of XO1b the de-correlation is significant but incomplete. The difference in maximum de-correlation achievable can be attributed to the systematic noise sources being strong functions of wavelength in the case of HD189733b whilst almost with constant weighting (${a}_{kl}$ in equation \ref{intro2}) in the case of XO1b. 

Whenever systematics have constant weighting per channel observed ($x_{k}$) and/or time, it becomes very difficult for PCA or ICA based approaches to de-correlate the signal from the systematics. Here auxiliary information of the instrument is needed to proceed with the de-correlation process. This is very well possible in the case of dedicated instruments such as Kepler, as these have been specifically designed with such high precision and stability measurements in mind \citep{borucki96,jenkins10}. For instruments that do not feature the calibration plan required to further de-correlate with instrument state parameters, the solution is far less obvious. 

We furthermore explored the de-correlation of eclipse signals observed consecutively rather than in parallel. We demonstrated, using Kepler data, that despite the formal violation of the `instantaneous mixing model', the proposed algorithm is able to retrieve the desired signal component with good accuracy. Such an application is particularly important for treating variability of the host-star which can significantly impair the quality of the final science result \citep[eg.][]{czesla09, boisse11, aigrain11,ballerini11}.

It is furthermore interesting to note that pre and post-processing steps (e.g. wavelets \citep{carter09}, Fourier based techniques \citep{waldmann11}, de-correlation using instrument state parameters \citep{swain08}),  do not break the instantaneous mixing model and can be run in conjunction with ICA methods. This makes independent component analysis a very powerful and versatile tool for non-parametric de-correlation of exoplanetary data sets.

\section{Conclusion}

In the light of searching and characterising ever smaller and fainter exoplanetary targets, the development of novel de-trending routines becomes increasingly critical. Based on the concepts of blind source deconvolution of instantaneously mixed signals, we have presented a first step towards non-parametric corrections and data filters that do not require additional information on the systematic noise of the instrument or stellar activity. Such algorithms have two important applications: 

1) For instruments that lack a calibration plan at the accuracy of 10$^{-4}$ in flux variation, which is required for spectroscopy of exoplanetary atmospheres, the spectroscopic signatures become inherently entangled and dependent on the method used to correct instrument and other systematics in the data. The de-correlation of spectroscopic data was demonstrated using two HST/NICMOS data sets. 

2) Detections of faint exoplanetary eclipses are often made difficult by time-correlated activity of the host star. We demonstrated, using a single Kepler time series, that much of the stellar variability can be removed in time series that span several exoplanetary eclipse events. 

The algorithm proposed is a powerful tool for lightcurve de-trending, which can be used by its own or in conjunction with any other type of data filtering or cleaning technique.  This becomes an invaluable advantage for data analysis when the instrument's response function is unknown or poorly characterised.

\acknowledgments

I.P.W. would like to thank Prof. E. Feigelson, the referee, Dr. G. Tinetti, Dr. F. Abdalla and Dr. S. Fossey for comments and suggestions that helped to greatly improve this paper. I.P.W. is supported by an STFC Studentship.

\appendix

This appendix provides some additional notes to the methods employed in this paper. For a more in-depth discussion of the topics presented here, please refer to the cited publications. 

\section{Uncorrelatedness, orthogonality and independence of Gaussian and non-Gaussian signals}
\label{appendix:independent}

In Gaussian statistics, our probability densities are fully defined by the first and second statistical moments, i.e. their means and covariances.
Two random vectors, $\bf{s_{l}}$ and $\bf{s_{l+1}}$, are said to be uncorrelated when their covariance (${\bf C}_{\bf{s}_l, \bf{s}_{l+1}}$) is zero:

\begin{eqnarray}
\label{cov}
\bf{C}_{{s}_l, {s}_{l+1}} &=&     E[({ s_{l}} - E[{s_{l}} ]) ({s_{l+1}} - E[{ s_{l+1}}])]  \\\nonumber
&=& E[{s_l}, { s_{l+1}}] - E[{ s_l}]E[{ s_{l+1}}] = 0
\end{eqnarray}

\noindent where $E[{s_{l}}]$ is the expectation value of ${s_{l}}$ which can be approximated by the mean in this case by

\begin{equation}
E[{ s_{l}}] \approx \frac{1}{M} \sum_{t=1}^{M}s_{l}(t)
\label{expectation}
\end{equation} 

\noindent with $M$ being the number of data points in the time series.

Furthermore, we define two random variables (${s_{l}}$ and ${s_{l+1}}$)  to be orthogonal, when both their  expectation values, in addition to their covariance are zero: 

\begin{equation}
\label{ortho}
\text{E}[{ s_{l}}] = \text{E}[{s_{l+1}}] = {\bf C}_{{s}_l, {s}_{l+1}} = 0
\end{equation}

We can always find an affine, linear transformation from a correlated set of variables to an orthogonal one.

Finally, our two random vectors ${s_{l}}$ and ${s_{l+1}}$ are independent from one another if and only if  the joined probability distribution $P({ s_{l}},{ s_{l+1}})$ of both signals  are factorizable into the product of their marginal pdfs, $P({ s_{l}})$ and $P({ s_{l+1}})$:

\begin{equation}
P({ s_{l}},{s_{l+1}}) = P({ s_{l}})P({ s_{l+1}})
\label{independence}
\end{equation}

\noindent and satisfy the property 

\begin{equation}
\text{E}[g({s_{l}} ) h(  s_{l+1} )] = \text{E}[ g( {  s_{l}} )] \text{E}[ h ({ s_{l+1}} )] 
\label{independence2}
\end{equation}

\noindent where $g({s}_{l})$ and $h({s}_{l+1})$ are absolutely integrable functions of ${s}_{l}$ and $s_{l+1}$ respectively. From the definition of independence in equation \ref{independence2}, we obtain the definition of uncorrelatedness (equation \ref{ortho}) in the special case where both ${ s_{l}}$ and ${ s_{l+1}}$ are linear and are only defined by their covariances (i.e. no higher order statistical moments) \citep{hyvarinen00, icabook, riley02}. 
In other words, uncorrelatedness is a special case of independence. Uncorrelated Gaussian random variables are always also independent and the definitions of uncorrelatedness, orthogonality (for zero mean) and statistical independence become identical.

\section{Preprocessing}

The covariance matrix of $\bf{X}$, $\bf{C}_{x}$, is given by $\bf{C}_{x} = \bf{E}\bf{D}\bf{E}^{T}$, where $\bf{E}$ is the matrix of eigenvectors and $\bf{D}$ the diagonal matrix of eigenvalues, $\textbf{D} = diag(d_{1}, d_{2},..., d_{n})$. Using principal component analysis (PCA), we compute \textbf{E} and \textbf{D} and the whitening matrix is hence the inverse square root covariance matrix $\bf{C_{x}^{-1/2}}$ is then given by equation \ref{Vwhite} \citep{icabook, pcabook}.

\begin{equation}
\bf{\tilde{X}} = \bf{C_{x}^{-1/2}}(\bf{X} - \bf{\bar{X}}) = \bf{\tilde{A}}\bf{S}
\label{whitening2}
\end{equation}

\begin{equation}
\bf{C_{x}^{-1/2}} = \bf{E} \bf{D}^{-1/2}\bf{E}^{T}
\label{Vwhite}
\end{equation}

\noindent where $\bf{\tilde{W}} \overset{\triangle}{=} \bf{\tilde{A}}^{-1}$ and is the de-mixing matrix of the whitened observed signals $\bf{\tilde{X}}$.

\section{Blind source separation}
\label{bss}

At the heart of the algorithm lies the blind-source separation routine. To attain the demixing matrix $\bf{\tilde{W}}$, many different types and varieties of algorithms are being used in the literature. Here we will use the 'Multi-COMBI' algorithm developed by \citet{tichavsky06} combining a fixed point high-order ICA algorithm to separate non-Gaussian sources with a second-order statistics blind-source-separation (BSS) algorithm for separating auto-regressive (AR) sources. We will briefly outline these algorithms and explain how it is applied to the whitened data $\bf{\tilde{X}}$ obtained in section \ref{pca}.


\subsection{Parallel FastICA and EFICA}
\label{efica}

In section~\ref{icadef} we briefly outlined the measures of non-Gaussianity, negentropy (equation~\ref{negentropy}),  used throughout this paper and stated that negentropy can be approximated via the kurtosis of the random vector $\bf{y}$ or via the use of contrast functions, equation~\ref{negentropy2} \& \ref{nonlin1}. In section~\ref{fastica} we showed stated the iteration scheme to obtain a single independent component (IC) at a time. This is called a deflationary algorithm where the computed IC is subtracted from the data before the second IC is computed. Such an iteration scheme has the property of finding the ICs in the order of decreasing non-Gaussianity. However, the main drawback of a serial computation of ICs is that estimation errors in the first ICs propagate in the extraction of later ICs via the orthogonalization step. This effect is cumulative and may signficantly impair weaker ICs. This predicament can be circumvented by estimating all ICs in parallel. 

Similar to the single unit iteration, the whitened demixing matrix $\bf{\tilde{W}}$ is at its most mutually independent when the projection $\bf{Y} = \bf{\tilde{W}}^{T}\bf{\tilde{X}}$ is at its most non-Gaussian. The FastICA fixed point iteration step is then given by 

\begin{equation}
\bf{\tilde{W}}^{+} \leftarrow g(\bf{\tilde{W}}\bf{\tilde{X}})\bf{\tilde{X}}^{T} - \text{diag}[g' (\bf{\tilde{W}}\bf{\tilde{X}})\bf{1}_{N}]\bf{\tilde{W}} 
\label{ica1}
\end{equation}

\noindent where $\bf{\tilde{W}}^{+}$ is the unnormalised next iteration of $\bf{\tilde{W}}$, $\bf{1}_{N}$ is an N~x~1 vector of 1's and $g(.)$ and $g'(.)$ are the first and second order derivatives of the nonlinear function $G(.)$:

 \begin{align}
\label{nonlin2}
g_{1}(y) &= \text{tanh}(a_{1}y) \\\nonumber
g_{2}(y) &= y \text{exp}(-y^{2}/2) \\\nonumber
g_{3}(y) &= y^{3} 
\end{align}

\noindent This is followed by a symmetric orthogonalisation step:

\begin{equation}
\bf{\tilde{W}} \leftarrow (\bf{\tilde{W}}^{+}\bf{\tilde{W}}^{+T})^{-1/2}\bf{\tilde{W}}^{+}
\label{ica2}
\end{equation}

\noindent Equations~\ref{ica1} \& \ref{ica2} are iterated until the result has converged. 

For a full derivation we refer you to \citet{hyvarinen99} and \citet{icabook}.
Whereas the convergence of the FastICA algorithm is often dependent on the non-linearity chosen by the user, the EFICA \citep{koldovsky06} algorithm employed here is a variant of the above iteration scheme and allows for different non-linearities to be assigned adaptively to different sources. \citet{koldovsky06} showed that EFICA is asymptotically efficient, ie. reaches the Cramer-Rao Lower Bound (CRLB) in an ideal case where the nonlinearity $G(.)$ equals the score function. 

To assert a good degree of separation, we can define $\bf{G}$ as the gain matrix. For a perfectly estimated de-mixing matrix, $\bf{W}$, the gain matrix is equal to its identity matrix  

\begin{equation}
\textbf{G} = \textbf{W}\textbf{A} = \textbf{I} 
\label{gainmatrix}
\end{equation}

In signal processing, the performance of blind-source separation algorithms is usually measured by the interference over signal ratio matrix, $\bf{ISR}$ 

\begin{equation}
\textbf{ISR}_{kl} = \frac{\textbf{G}^{2}_{kl}}{\textbf{G}^{2}_{kk}}, ~ k,l = 1,2,...,d
\label{ISR}
\end{equation}

\noindent where $k$ and $l$ denote the observed and estimated sources. The ISR for and individual observed signal $k$ is given by

\begin{equation}
\textbf{isr}_{k} = \frac{\sum^{d}_{l=1,l \neq k} \textbf{G}^{2}_{kl}} {\textbf{G}^{2}_{kk}}, ~ k = 1,2,...,d
\label{isrk}
\end{equation}

However, the original mixing matrix, \textbf{A}, is not generally known for real data sets and equations \ref{ISR} \& \ref{isrk} are only useful in the case of simulations. \citet{tichavsky06} have shown that the whole \textbf{ISR} matrix for the EFICA algorithm can be approximated by

\begin{equation}
\label{ISRef}
\textbf{ISR}^{EF}_{kl} \simeq \frac{1}{N} \frac{\gamma_{k}(\gamma_{l}+\tau_{l}^{2})}{\tau^{2}_{l}\gamma_{k} + \tau^{2}_{k}(\gamma_{l} + \tau^{2}_{l})}
\end{equation}

\begin{align}
\gamma_{k} &= \beta_{k} - \mu_{k}^{2}  \\\nonumber 
\mu_{k} &= E[\hat{s}_{k} g_{k}(\hat{s}_{k})] \\\nonumber 
\tau_{k} &= |\mu_{k} - \rho_{k}| \\ \nonumber 
\rho_{k} &= E[g_{k}'(\hat{s}_{k}] \\\nonumber 
\beta_{k} &= E[g_{k}^{2} (\hat{s}_{k}]
\label{ISRef}
\end{align}

\noindent where  $\hat{\bf{s}}_{k}$ and $\hat{\bf{s}}_{l}$ are the k'th and l'th observed and estimated signals of \textbf{S} in equation \ref{timeseries3}, $g_{k}(.)$ and $g'_{k}(.)$ the first and second derivative of $G(.)$ for signal $k$ and $N$ is the number of signals estimated. 
Here it should be mentioned that, of course, the true realisation of each \textbf{ISR} component is unknown and a mean-\textbf{ISR} is computed leading to the best 'on average' separation of the signals.

\subsection{WASOBI}
\label{wasobi}

Whilst EFICA is optimised for the separation of instantaneously mixed, non-Gaussian sources, second-order statistics BSS algorithms rely on time-structure in the sources' correlation function to estimate $\bf{\tilde{W}}$. A variety of algorithms exist in the literature, here we use a derivative of the popular SOBI algorithm \citep{belouchrani97}, WASOBI \citep{yeredor00, tichavsky06b} to separate Gaussian auto-regressive (AR) sources in the input data $\bf{\tilde{X}}$.
Here, the blind source separation follows the same linear model as in equation \ref{timeseries3} and the mixing matrix $\bf{\tilde{A}}$ is estimated by a joint diagonalisation of the signals' autocorrelation matrices. The unknown correlation matrices of the observed signals for a given lag $\tau$, $\textbf{R}_{x}[\tau]$

\begin{equation}
\textbf{R}_{x}[\tau] \overset{\triangle}{=} \frac{1}{N} \sum_{n=1}^{N} \textbf{x}[n] \textbf{x}^{T}[n+\tau], ~~ \tau = 0, ..., M-1
\label{wasobi1}
\end{equation}

\noindent satisfies the relation 

\begin{equation}
\textbf{R}_{x}[\tau] = \bf{\tilde{A}R}_{s}[\tau]\bf{\tilde{A}}^{T},~~~~ \forall \tau
\label{wasobi2}
\end{equation}

\noindent where $\textbf{R}_{s}[\tau] \overset{\triangle}{=} E[ \textbf{s}[n] \textbf{s}^{T}[n+\tau]]$ are the source signals' diagonalised correlation matrices \citep{yeredor00}. Hence, if the correlation matrices are diagonal, ie. the off-diagonal components are zero, the separated signals can be said to be independent from each other. The SOBI \& WASOBI algorithms estimate $\bf{\tilde{A}}$ as the joint diagonoliser of a set of correlation matrices. Similar to the EFICA code, we can define an asymptotic estimate of the \textbf{ISR} matrix

\begin{equation}
\textbf{ISR}_{kl}^{WA} \simeq \frac{1}{N} \frac{\phi_{kl}}{1- \phi_{kl}\phi_{lk}} \frac{\sigma_{k}^{2} R_{l}[0]}{\sigma_{l}^{2}R_{k}[0]}
\label{ISRwa}
\end{equation}

\begin{equation}
\phi_{kl} \overset{\triangle}{=} \frac{1}{\sigma^{2}_{k}} \sum_{i,j =0}^{M-1} a_{il} a_{jl} R_{k}[i-j]
\label{phi}
\end{equation}

\noindent where $k$ and $l$ denote the observed and the estimated sources, $\{R_{k}[\tau]\}^{M-1}_{\tau=0}$ is the covariance sequence of the $k$-th source, $\sigma^{2}_{k}$ is the variance of the source and $\{a_{il}\}^{M-1}_{i=0}$ are the auto-regression coefficients of the $l$-th source \citep{tichavsky06b}.

\subsection{Multi-COMBI}
\label{mcombi}

The algorithms introduced above are highly complementary to each other. Whilst EFICA has an asymptotically efficient performance in separating non-Gaussian instantaneous mixtures, WASOBI is asymptotically efficient in separating Gaussian time-correlated signals. Both these properties are necessary since a real data set will have both of the aforementioned properties and its components would hence not be optimally de-mixed if one would only employ one type of algorithm. 
MULTI-COMBI \citep{tichavsky06} uses a clustering technique in which both algorithms are run on the set of unseparated sources $\bf{\tilde{X}}$ and their interference over signal matrices, $\bf{ISR}^{EF}$ and $\bf{ISR}^{WA}$, are estimated. The signals are then clustered depending on whether their specific $\bf{ISR_{kl}}$ is lower for the EFICA or WASOBI case. Then, the process is repeated until all clusters are singeltons, ie. only contain one signal per cluster, and the signals are hence optimally separated. 

\section{Convergence check}

From the MULTI-COMBI algorithm, we obtain the estimated signal matrix $\bf{\hat{S}}$, an overall \textbf{ISR} matrix as well as final $\bf{ISR}^{EF}$ and $\bf{ISR}^{WA}$. Since the algorithms used here use fixed-point convergence techniques, the problem of non-repeatability of the separation process is less than for neural network based approaches. However, it is common sense to check the stability of the result obtained and to estimate the error on $\bf{\hat{S}}$. 

In order to estimate the stability of the convergence, we perturb the unknown mixing matrix \textbf{A} with a random and known mixing matrix $\bf{P}$ to give a new mixing matrix $\bf{A}_{2} = \bf{P}\bf{A}$ and equation \ref{timeseries3} becomes: $\bf{X} = \bf{P}\bf{A}\bf{S} = \bf{A}_{2}S$. This is equivalent to multiplying the whitened signal $\bf{\tilde{X}}$ with \textbf{P}

\begin{equation}
\bf{\tilde{X}}_{2} = \bf{P} \bf{\tilde{X}} = \bf{P}\bf{C_{x}^{-1/2}} (\bf{X} - \bf{\bar{X}}) = \bf{\tilde{A}}_{2}\bf{S}
\label{convtest}
\end{equation}

We re-run the separation step and estimate $\bf{A}_{2}$. Since \textbf{P} is known, we can reconstruct the original mixing-matrix and compare it with the new result. In the scope of an automated algorithm, the sum of all terms of $\bf{ISR}_{A}$ is compared to the sum of $\bf{ISR}_{A2}$ and the result is reported.  

To identify the stochastic nature of the retrieval we furthermore re-run the separation step with the same whitened signal, $\bf{\tilde{X}}$, akin to a Monte Carlo simulation. We perform $i$ realisations (where $i = 10-100$ typically) and use the de-mixing matrices $\bf{\tilde{W}}_{i}$ to construct mean noise models later on. This way, we propagate the signal separation error to the model-fitting in a coherent manner.

\section{Signal separation}

In order to identify the non-white (i.e. systematic) signals in our estimated signal matrix $\bf{\hat{S}}$, we use the Ljung-Box portmanteau test \citep{brockwell06}.
The test statistic, usually denoted by $Q$, is defined by summing the normalised autocorrelations of the individual time series, $\bf{\hat{s}}_{l}$ over a range of lags:

\begin{equation}
Q = n(n + 2) \sum_{\tau=1}^{m} \frac{\hat{\rho}^{2}_{\tau}}{m-\tau}
\label{ljung1}
\end{equation}

\noindent where $\hat{\rho}^{2}_{\tau}$ is the autocorrelation at lag $\tau$ and $m$ is the number of observations in the time series. The hypothesis of the time series being solely white noise is rejected if $Q$ is bigger than a pre-specified fraction of the chi-squared distribution

\begin{equation}
Q > \chi^{2}_{1-\alpha,h}
\label{ljung2}
\end{equation}

\noindent where $\chi^{2}_{1-\alpha,h}$ is the $\alpha$-quantile of the chi-squared distribution with $h$ degrees of freedom \citep{brockwell06}. Here we take $\alpha = 0.05$.

\end{document}